\documentclass[traditabstract]{aa} 
\usepackage{graphicx}
\usepackage{txfonts}
\usepackage{natbib}
\bibpunct{(}{)}{;}{a}{}{,}
\usepackage{float}
\usepackage{calc}
\usepackage{textcomp}
\usepackage{placeins}
\usepackage{color}
\usepackage{rotating} 
\usepackage{lscape}
\usepackage{url}
\usepackage{subfig}
\usepackage[all]{draftcopy}
\usepackage{longtable}
\usepackage{array}
\usepackage{threeparttable}
\DeclareTextSymbol{\degre}{OT1}{23}
\newcounter{savedfootnote}

\begin{document}
   \title{Submillimetre Photometry of 323 Nearby Galaxies from the {\it Herschel}\thanks{{\it Herschel} is an ESA space observatory with science instruments provided by European-led Principal Investigator consortia and with important participation from NASA.} Reference Survey. }

\author{L. Ciesla\inst{1},
	  A.Boselli\inst{1},
	  M. W. L. Smith\inst{2},	  
	  G. J. Bendo\inst{3},
	  L. Cortese\inst{4},
	  S. Eales\inst{2},
	  S. Bianchi\inst{5},
	  M. Boquien\inst{1},
	  V. Buat\inst{1},
	  J. Davies\inst{2},
	  M. Pohlen\inst{2},
	  S. Zibetti\inst{5,6},
	  M. Baes\inst{7},
	  A. Cooray\inst{8,9},
	  I. de Looze\inst{7},
	  S. di Serego Alighieri\inst{5},
	  M. Galametz\inst{10},
	  H. L. Gomez\inst{2},
	  V. Lebouteiller\inst{11},
	  S. C. Madden\inst{11},
	  C. Pappalardo\inst{5},
	  A. Remy\inst{11},
	  L. Spinoglio\inst{12},
	  M. Vaccari\inst{13,14},
	  R. Auld\inst{2},
	  D. L. Clements\inst{15}.
         }

\institute{	
		Laboratoire d'Astrophysique de Marseille - LAM, Universit\'e d'Aix-Marseille \& CNRS, UMR7326, 38 rue F. Joliot-Curie, 13388 Marseille Cedex 13, France               	
         \and 
		School of Physics and Astronomy, Cardiff University, Queens Buildings The Parade, Cardiff CF24 3AA, UK
	\and
		UK ALMA Regional Centre Node, Jodrell Bank Centre for Astrophysics, School of Physics and Astronomy, University of Manchester, Oxford Road, Manchester M13 9PL, United Kingdom
	 \and
	 	European Southern Observatory, Karl Schwarzschild Str. 2, 85748 Garching bei Muenchen, Germany
	 \and
		INAF-Osservatorio Astrofisico di Arcetri, Largo Enrico Fermi 5, 50125 Firenze, Italy
	 \and 
		Dark Cosmology Centre, Niels Bohr Institute University of Copenhagen, Juliane Maries Vej 30, DK-2100 Copenhagen, Denmark
	\and
		Sterrenkundig Observatorium, Universiteit Gent, Krijgslaan 281 S9, 9000, Gent, Belgium
	\and
		Department of Physics \& Astronomy, University of California, Irvine, CA 92697, USA 10
	\and	
		California Institute of Technology, 1200 E. California Blvd, Pasadena, CA 91125, USA
	 \and 
		Institute of Astronomy, University of Cambridge, Madingley Road, Cambridge CB3 0HA
	 \and 
		CEA/DSM/IRFU/Service d'Astrophysique, CEA, Saclay, Orme des Merisiers, Batiment 709, F-91191 Gif-sur-Yvette, France
	\and
		Istituto di Fisica dello Spazio Interplanetario, INAF, Via Fosso del Cavaliere 100, I-00133 Roma, Italy
	\and
		Dipartimento di Astronomia, Universit\`a di Padova, vicolo Osservatorio, 3, 35122 Padova, Italy	
	\and
		Astrophysics Group, Physics Department, University of the Western Cape, Private Bag X17, 7535, Bellville, Cape Town, South Africa	
	\and
		Astrophysics Group, Imperial College, Blackett Laboratory, PrinceConsort Road, London SW7 2AZ, UK
}

   \date{Received; accepted}

  \abstract
  {
   The \textit{Herschel} Reference Survey (HRS) is a guaranteed time \textit{Herschel} key project aimed at studying the physical properties of the interstellar medium in galaxies of the nearby universe.	
   This volume limited, K-band selected sample is composed of galaxies spanning the whole range of morphological types (from ellipticals to late-type spirals) and environments (from the field to the centre of the Virgo Cluster). 
   We present flux density measurements of the whole sample of 323 galaxies of the HRS in the three bands of the Spectral and Photometric Imaging Receiver (SPIRE),  at 250~$\mu$m, 350~$\mu$m and 500~$\mu$m.   
   Aperture photometry is performed on extended galaxies and point spread function (PSF) fitting on timeline data for unresolved objects; we carefully estimate errors and upper limits.
   The flux densities are found to be in good agreement with those of the HeViCS and KINGFISH key projects in all SPIRE bands, and of the \textit{Planck} consortium at 350~$\mu$m and 550~$\mu$m, for the galaxies in common.
   This submillimetre catalogue of nearby galaxies is a benchmark for the study of the dust properties in the local universe, giving the zero redshift reference for any cosmological survey. 
   }

   \keywords{Galaxies: ISM; Infrared: galaxies; Surveys; Catalogs }
  
   \authorrunning{Ciesla et al.}
   \titlerunning{The SPIRE photometry of the \textit{Herschel} Reference Survey}

   \maketitle

\section{\label{intro}Introduction}

	Dust grains in the interstellar medium (ISM) of galaxies profoundly affect our view of these systems by absorbing the ultra-violet (UV) and optical stellar emission and re-emitting it in the infrared, from $\sim$5~$\mu$m to $\sim$1~mm.
	Dust is produced by the aggregation of metals injected into the interstellar medium by massive stars, through stellar winds \citep{Hofner09,Gomez10b}, supernovae \citep{Clayton97,Bianchi07,Matsuura11,Gomez12}, or less massive stars in their final evolution stages, such as asymptotic giant branch stars \citep{Gehrz89,Dwek98,Galliano08}. 
	Generally intermixed with gas in the ISM, dust is thus a good tracer of the cold molecular and atomic phases of the ISM and contains a significant fraction of metals.
	Dust plays an important role in the interstellar medium as it acts as a catalyst  in the transformation process of the atomic to molecular hydrogen, shields the UV radiation field preventing the dissociation of molecular clouds, and contributes to the cooling and heating of the ISM in photodissociation regions \citep{Wolfire95}.
	The 5-70~$\mu$m spectral range corresponds to the emission of the hot dust generally associated with star formation whereas at longer wavelengths, up to $\sim$1mm, the submillimetre emission is generally dominated by the emission from the cold dust \citep[e.g.][]{Bendo10,Bendo12,Boquien11}.
	\textit{IRAS} (Neugebauer et al. 1984), \textit{COBE} (1989), \textit{ISO} (Kessler et al. 1996), \textit{Spitzer} (Werner et al. 2004), and \textit{AKARI} \citep{Murakami07} allowed us to study the emission of the dust up to 240~$\mu$m. 
	However, most of the cold dust emission is drowned by warm dust at wavelengths shorter than 240~$\mu$m. 
	An accurate determination of dust masses requires submillimetre data \citep{DevereuxYoung90,Gordon10,Galametz11,Bendo12}.
	Ground-based facilities, such as SCUBA on \textit{JCMT} \citep{Holland99}, reveal the submillimetre domain but observations of large samples of normal galaxies such as ours is still prohibitive due to the long integration times needed for these instruments.
	The \textit{Herschel} Space Observatory (Pilbratt et al. 2010), launched in May 2009, opens a new window on the far-infrared/submillimetre spectral domain (55 to 672~$\mu$m) and allows us to probe the cold dust component in large numbers of nearby galaxies.
	
	To characterize the dust properties in the local universe, the Spectral and Photometric Imaging Receiver (SPIRE) \citep{Griffin10} Local Galaxies Working Group (SAG 2) has selected 323 galaxies to be observed as part of the \textit{Herschel} Reference Survey (HRS) \citep{Boselli10a}.
	The HRS is a guaranteed time key project and a benchmark study of dust in the nearby universe. 
	The goals of the survey are to investigate (i) the dust content of galaxies as a function of Hubble type, stellar mass and environment, (ii) the connection between the dust content and composition and the other phases of the interstellar medium, and (iii) the origin and evolution of dust in galaxies. 
	The HRS spans the whole range of morphological types including ellipticals to late-type spirals, with a few irregular dwarf galaxies, and environments (from relatively isolated field galaxies to members of the core of the Virgo Cluster). 
	The sample is ideally defined also because of the availability of a large set of ancillary data.
	Multiwavelength data, from the literature, are available for about 90\% of HRS galaxies from \textit{IRAS} \citep{Sanders03, Moshir90, ThuanSauvage92, Soifer89, Young96}; optical and near-infrared from SDSS \citep{SDSS09} and 2MASS \citep{Jarrett03}, and radio from NVSS \citep{Condon98} Êand FIRST \citep{Becker95}.
	Most of these data are available on NED and GOLDMine \citep{Gavazzi03}.
	Some of them are already released or currently analysed by our team such as UV data from \textit{GALEX} \citep[][Cortese et al., submitted]{Boselli11b}, \textit{Spitzer}/IRAC (Ciesla et al., in prep), and \textit{Spitzer}/MIPS \citep{Bendo12b}. 
	H$\alpha$ imaging (Boselli et al., in prep), CO(J=1-0) spectroscopy (Boselli et al., in prep), optical integrated spectroscopy (Boselli et al., submitted), CO(J=3-2) \textit{JCMT} mapping (Smith et al. in prep), and gas metallicities (Hughes et al. 2012, submitted) will be soon available.

	The first scientific results, based on a subsample of HRS data obtained during the Science Demonstration Phase (SDP) data, were presented in the A\&A \textit{Herschel} Special Issue (2010).
	Statistical studies based on this data set investigate the far-infrared/submillimetre colours \citep{Boselli12}, the dust scaling relations as a function of environment and galaxy type \citep{Cortese12}, and the properties of the early-type galaxies in the sample \citep{Smith12}. 
	A preliminary analysis of the spectral energy distributions (SEDs) was done as part of the SDP \citep{Boselli10b} while the complete analysis and modelling of the SEDs of the whole sample is in preparation (Ciesla et al., in prep).
	Thanks to the high angular resolution of \textit{Herschel} ($\sim$18$\arcsec$ at 250~$\mu$m, leading to a resolution of a few kpc at 20~Mpc), studies of large resolved HRS galaxies (with angular sizes between 2$\arcmin$ and 10$\arcmin$)  within kiloparsec-sized subregions are also possible \citep{Bendo12,Boquien12}.
	
	The aim of this paper is to present the HRS catalogue of the flux densities at 250, 350 and 500~$\mu$m.
	The paper is organised as follows. 
	In Section~\ref{sample}, we briefly describe the HRS sample.
	Section~\ref{obsdata} gives the description of the \textit{Herschel}/SPIRE observations and data reduction. 
	Section~\ref{fluxex} details the techniques used for the flux extraction.
	In Section~\ref{fluxdens}, we provide flux densities of the whole sample and in Section~\ref{comp}, we compare our results to those available in the literature.

\section{\label{sample}The sample}

The HRS galaxies are selected according to several criteria fully described in \cite{Boselli10a}.
The HRS is a volume limited sample composed of galaxies at a distance between 15 and 25 Mpc.
The galaxies are selected according to their K band magnitude, whose luminosity is a proxy for the total stellar mass \citep{Gavazzi96}. 
Based on the optical extinction studies and far-infrared observations at wavelength shorter than 200~$\mu$m, we expect late-type galaxies to have a larger content of dust than early-types \citep{Sauvage94}.
Thus, two different $K_{mag}$ limits have been adopted: $K_{mag}\leq$12 for late-types and $K_{mag}\leq$8.7 for early-types (in Vega magnitudes).
Finally, to limit any contamination from Galactic cirrus, we selected galaxies at high Galactic latitude ($b > + 55\deg$) and with low Galactic extinction regions \citep[$A_{B} < 0.2$][]{Schlegel98}.
The final sample contains 323\footnote{With respect to the original sample given in \cite{Boselli10a}, the galaxy HRS~228 should be removed from the complete sample because its updated redshift on NED indicates it as a background object.}  galaxies, 62 early-types and 261 late-types.

\section{\label{obsdata}\textit{Herschel}/SPIRE observations and data reduction}

	\subsection{\label{obs}Observations}	
	We observed the 323 HRS galaxies with SPIRE \citep{Griffin10} in three wide bands at 250, 350 and 500~$\mu$m. 
	In order not to duplicate \textit{Herschel} observations, 79 galaxies out of 323 were observed as part of the open time key project the \textit{Herschel} Virgo Cluster Survey (HeViCS, Davies et al. 2010).
	
	For the 239 galaxies outside the Virgo cluster plus 4 Virgo galaxies observed during the Science Demonstration Phase, the observations are carried out using the SPIRE scan-map mode with a nominal scan speed of 30\arcsec s$^{-1}$.
	The sizes of the images depend on the optical extent of the targets and have been chosen to cover 1.5 times the optical diameter, $D_{25}$\footnote{The diameter at 25 mag arcsec$^{-2}$.}, of the galaxies, which is the full area over which the infrared emission is expected. 
	Previous observations of spiral discs indeed indicated that the infrared emission of late-type galaxies can be more extended than the optical disc \citep{Bianchi00}.
	For galaxies with an optical diameter smaller than $\sim 3\arcmin$, the small scan-map mode is used to provide a homogeneous coverage of a circular area of $\sim 5\arcmin$ diameter.
	For galaxies with an optical diameter larger than $\sim 3\arcmin$, the large scan-map mode is used to cover, at least, $1.5\times D_{25}$.
	The resulting sizes of the maps are thus $8\arcmin\times8\arcmin$, $12\arcmin\times12\arcmin$, and $16\arcmin\times16\arcmin$, depending on the optical size of the target.
	As early-type galaxies are known to contain less dust than late-types \citep{Ferrarese06}, longer integration times were used on this subsample \citep{Smith12}.
	For late-types, 3 pairs of cross-linked scan maps are made, while 8 pairs for early-types.

	HeViCS covers $55\deg^{2}$ at full depth ($84\deg^{2}$ in total), of the centre of the Virgo cluster at five wavelengths, from 100 to 500~$\mu$m, using the PACS/SPIRE parallel mode, down to the confusion limit in the SPIRE bands \citep{Davies12}.
	For the 79 HRS galaxies observed by HeViCS, the PACS/SPIRE parallel mode scan map is used with a scan speed of 60\arcsec\ s$^{-1}$ done with 4 pairs of perpendicular scans.
	Regions around each of the 79 galaxies are cut off from the large fields to perform the aperture photometry.
	They are large enough to provide a good estimate of the background emission.

	\subsection{\label{data}Data reduction}	
	
	The complete description of the data reduction and map-making procedures will be presented in a dedicated paper by Smith et al. (in prep).   
	Here, we just give a brief summary of the different steps carried out within the \textit{Herschel} Interactive Pipeline Environment software \citep[HIPE;][]{Ott11}.
	The SPIRE data are processed up to Level-1, the level where the pointed photometre timelines are derived, with a script adapted from the official scan map pipeline \citep{Griffin09,Dowell10}.
	The only difference to the Level-1 product is that we use the optimised deglitcher setting available for the observing mode.
	The typical SPIRE pipeline performs the following data processing steps on the timelines:	
	\begin{enumerate}
		\item Application of the glitch removal procedure to delete the cosmic rays that affect all detectors in an individual array, and then application of the wavelet glitch removal of cosmic rays from individual detectors (\textsc{WaveletDeglitcher}) for HRS data. For HeViCS data, \textsc{the sigmaKappaDeglitcher} is used.
		\item Application of an electrical low pass filter response correction, to correct for the delay in the data coming out of the electronics; this matches the detector timelines to the astrometric pointing timelines.
		\item Reapplication of the wavelet glitch removal for the HRS data; this additional step improves the removal of all glitches.
		\item Application of an additional time response correction. 
		\item Flux calibration, which includes nonlinearity corrections.
		\item Removal of the temperature drift where all bolometers are brought to the same level using a custom method called BriGAdE (Smith et al. in prep).
		\item Corrections of the bolometer time response, which adjusts the bolometer detector timelines to account for the fact that the bolometers do not respond instantaneously to signal.		
		\item Creation of the final maps using the naive map maker included in the standard pipeline.
	\end{enumerate}
	The pipeline also performs some steps related to associating the astrometry with the bolometer timeline data and performs some minor time corrections before the mapmaking step.

	\begin{table*}
	\centering
	\begin{threeparttable}[b]
	\caption{Summary of the properties of SPIRE image data.}
	\begin{tabular}{l c c c}
  	\hline\hline
  	  											& 250~$\mu$m  	& 350~$\mu$m 			& 500~$\mu$m\\
	\hline											
  	Pixel size ($pixsize_{\rm \lambda}$)				& 6\arcsec			& 8\arcsec				& 12\arcsec  \\ 
	Map FWHM									& 18.2\arcsec		& 24.5\arcsec			& 36.0\arcsec \\
  	Beam area ($beam_{\rm \lambda}$)			 	& 423 arcsec$^{2}$	& 751 arcsec$^{2}$ 	& 1587 arcsec$^{2}$ \\ 
	\hline
	Correction for extension ($K_4$)				& $0.98279$		& $0.98344$			& $0.97099$  \\ 
	Correction for updated calibration				& 1.					& $1.0067$ 			& 1.  \\ 	
	Total correction	 ($corr_{\rm \lambda}$)			& $0.98279$		& $0.99003$  			& $0.97099$  \\ 
	\hline
	\label{corr}
	\end{tabular}
	\end{threeparttable}
	\end{table*}

  	The pixel sizes and FWHM values of the final maps are provided in Table~\ref{corr}.
	By default, the pipeline applies a correction to the maps, called $K_4p$, that converts the flux densities weighted by the relative spectral response function (RSRF) in monochromatic flux densities, corresponding to spectra where $\nu S_\nu$ is constant.
	In doing this, the pipeline considers all sources as point-like objects.
	However, the RSRF changes for extended sources, so we need to divide the data by the $K_4p$ for point sources, automatically applied by the pipeline, and apply the $K_4e$ for extended sources instead.
	For such extended sources, the resulting correction is thus ${K_4e}/{K_4p}$, and their values are provided by the SPIRE Observer Manual\footnote{\url{http://herschel.esac.esa.int/Docs/SPIRE/html/spire_om.html}}.
	There are, in our sample, galaxies which are almost point-like that would not need this $K_4$\footnote{For simplicity, here we define $K_4$=${K_4e}/{K_4p}$, for the exact definition refer to the SPIRE Observer Manual} correction.
	Defining if a source is barely resolved can be subjective, thus for clarity, we defined two groups: one for point-like sources, one for extended sources using a quantitive criterion (see Section~\ref{pl}).
	We apply this $K_4$ correction for extended sources on all resolved galaxies of the HRS sample.
	
	To update the maps product for the present work to the latest calibration (HIPE v8, SPIRE calibration tree v8.1),  we multiply all 350~$\mu$m measurements by 1.0067, the most recent flux calibration (SPIRE photometry cookbook\footnote{\url{http://herschel.esac.esa.int/twiki/pub/Public/SpireCalibrationWeb/SPIREPhotometryCookbook_jul2011_2.pdf}}, Bendo et al. 2011).
	These corrections are listed in Table~\ref{corr}, we define the total correction $corr_{\rm \lambda}$ as the combination of correction for extension ($K_4$) and the correction for updated calibration.
	
		\subsection{Colour corrections}

	The SPIRE flux calibration assumes that the sources have a spectrum with $\nu S_{\nu}$ constant across the filter. 
	This assumption does not correspond to the SEDs of the objects observed \citep{Boselli10b}.
	The observed SEDs of the target galaxies are close to a modified black body with a spectral index ranging from $\beta$ $\sim$ 1.0 to 2.0 \citep{Boselli12}.
	Tuned colour corrections could thus be required.
	To quantify these colour corrections, we assume that the far-infrared spectrum can be farely well represented by a modified black body with a spectral index $\beta$ of 1.5 or 2.0
	Table~\ref{colcorr}  lists the colour corrections that should be applied for different sets of $\beta$ and $T$.
	They were obtained by integrating a modified blackbody of given $\beta$ and $T$ over the SPIRE spectral response function for each band. 
	For extended sources, the spectral response functions have been weighted by $\lambda^2$ (SPIRE Observer Manual) thus resulting in different sets of colour corrections for extended sources with respect to point sources. 	
	Indeed, for feedhorn bolometers in general and for the detectors in SPIRE specifically, the relative spectral responsivity function (RSRF) changes between point-like and extended sources as explained in Section 5.2.1 of the SPIRE Observers' Manual.  
	The colour corrections rely upon the integral of the product of the spectrum and the RSRF, then if the RSRF changes, the colour corrections will also change.
	We give the values for both point-like and extended sources.
	They are multiplicative corrections.
	Given the still poorly constrained shape of the SED of the target galaxies, these corrections listed in Table~\ref{colcorr} are not applied to the set of data given in Table~\ref{HRS_phot}.

	\begin{table*}[width=\columnwidth]
	\centering
	\caption{\label{colcorr} The colour corrections for the SPIRE data for extended and point-like sources.}
	\begin{tabular}{c c c c c c c c } 
	\hline
	\hline
	\multicolumn{2}{c}{} & \multicolumn{3}{c}{Extended} &  \multicolumn{3}{c}{Point like} \\  
	\multicolumn{2}{c}{} & 250~$\mu$m & 350~$\mu$m & 500~$\mu$m & 250~$\mu$m & 350~$\mu$m & 500~$\mu$m\\ 
	\hline
	$\beta=2$ 	& $T=10K$  	& 1.026 	& 1.025 	& 1.044 	& 1.023 & 0.995 & 0.959 \\ 
				& $T=15K $ 	& 1.019 	& 1.009 	& 1.021 	& 0.984 & 0.959 & 0.917 \\ 
				& $T=20K $ 	& 1.005 	& 0.996	& 1.007 	& 0.955 & 0.938 & 0.896 \\
				& $T=25K $	& 0.994 	& 0.988 	& 0.999 	& 0.937 & 0.925 & 0.884 \\	
				& $T=30K$ 	& 0.985 	& 0.983 	& 0.994 	& 0.924 & 0.917 & 0.876 \\
	\hline
	$\beta=1.5$  & $T=10K$ 	& 1.021 	& 1.025 	& 1.049 	& 1.026 & 1.004 & 0.977 \\ 
				& $T=15K$ 	& 1.022 	& 1.014 	& 1.031 	& 0.995 & 0.972 & 0.940 \\ 
				& $T=20K$ 	& 1.011 & 1.004 & 1.020 	& 0.970 & 0.953 & 0.920  \\ 
				& $T=25K$ 	& 1.002 & 0.998 & 1.014	& 0.953 & 0.942 & 0.909  \\
				& $T=30K$ 	& 0.995 & 0.993 & 1.009	& 0.941 & 0.934 & 0.902 \\	
				& $T=35K$ 	& 0.990 & 0.990 & 1.006	& 0.933 & 0.929 & 0.897  \\
				& $T=40K$ 	& 0.986 & 0.987 & 1.003 	& 0.926 & 0.925 & 0.893 \\
				& $T=45K$ 	& 0.983 & 0.985 & 1.001 	& 0.922 & 0.922 & 0.890 \\
				& $T=50K$ 	& 0.981 & 0.984 & 1.000 	& 0.918 & 0.919 & 0.888 \\
	\hline			
	\end{tabular}
	\tablefoot{ Multiplicative factors to apply to correct values for galaxies with spectral energy distributions well represented by a modified black body with a grain emissivity parameter $\beta$ and a temperature $T$ in the given ranges. Color corrections for $\beta=2$ and $T=15,20,25K$, are consistent with those given in \cite{Davies12}; in that paper, however, the $K_4$ correction is included in the color correction for extended sources, while here it is included in the fluxes.}\\
	\end{table*}


\section{\label{fluxex}Flux extraction}

	As we chose long integration times in order to reach the confusion limit, the images include faint background sources.
	Furthermore, despite a high Galactic latitude selection, some images show the presence of Galactic cirrus.
	Our sample contains large extended galaxies, point-like sources and non-detected galaxies.
	These technical aspects, related to the nature of the emitting source and of the sky background emission around it, prevent us from performing an automatic photometry. 
	Three methods are needed to accurately measure the flux for all of objects: a first method for point-like sources (PSF fitting on timeline data), a second for extended ones (aperture photometry) and the third for the determination of upper limits.

	\subsection{\label{pl}Point-like sources}	
	SPIRE is calibrated using a timeline-based PSF fitting approach.
	It is then possible to extract the flux densities of point-like sources directly from the timeline data using a PSF fitting method.
	The PSF fitter fits a two-dimensional Gaussian function to the signal and position timeline data.  
	For unresolved sources, the peak of the Gaussian function from the fit corresponds to the flux density of the source.  
	For more information, see Bendo et al. (2012, in prep).
	Based on tests of aperture photometry versus timeline-based PSF fitting for SPIRE data, there is significantly worse accuracy and precision for aperture photometry on unresolved sources.  
	Moreover, there are some systematic effects that are actually due to the mapping technique (related to where bolometers tend to cross over the unresolved sources).  
	Timeline-based PSF fitting actually avoids those biases for point sources.
	The SPIRE-ICC (Instrument Control Center)  strongly recommends the use of PSF fitting on timeline data for unresolved sources (SPIRE Observer's Manual\footnote{\url{http://herschel.esac.esa.int/Docs/SPIRE/html/spire_om.html}} Section 5.2.11).  
	
	To identify point-like sources and measure their flux densities, we proceed with the following method.		
	All of the images are inspected in order to make a list of point-like sources candidates.
	We run the timeline-based PSF fitter program and use the criterion given by the SPIRE photometry cookbook. 
	For a given band, if the FWHM of the resulting gaussian fitted to the timeline data is smaller than 20\arcsec, 29\arcsec and 37\arcsec at 250, 350 and 500~$\mu$m, respectively, then the source is considered as point-like.
	These limits for the FWHM were determined empirically by adding artificial randomly-placed sources to timeline data and then performing timeline-based PSF fits to those data.  
	The resulting distribution of FWHM indicates that 20\arcsec, 29\arcsec and 37\arcsec are acceptable upper limits for the typical FWHM that will be measured for sources.	
	According to this criterium, in the whole HRS sample, there are 10, 10 and 9 point-like sources at 250, 350 and 500~$\mu$m, respectively.
	As the timeline data are calibrated in Jy/beam, the timeline fits for these data give amplitudes that correspond to the flux densities of the sources.
	These will be the most accurate flux densities that can be measured for these sources, as the measurement technique matches the method applied to the primary and secondary sources used for SPIRE flux calibration.
	As the pipeline is optimised for point sources, their flux densities do not need to be corrected with the $K_4$ correction described in Section~\ref{data} and provided in Table~\ref{corr}. 
	However, the 350~$\mu$m measurements are corrected for the HIPE v8 updated calibration, thus they are multiplied by 1.0067.

	\subsection{\label{ext}Extended sources}
	
	The aperture photometry of extended sources is carried out using the DS9/Funtools program ``Funcnts''.
	This task performs a basic aperture photometry, summing all pixels within a defined elliptical region.
	The mean value of the background is calculated in a given annulus and then subtracted from the counts of the aperture.
	With ``Funcnts'', we can  extract the counts in elliptical regions adapted to match the shape of the galaxies.

\begin{figure*}
 		\includegraphics[width=\textwidth]{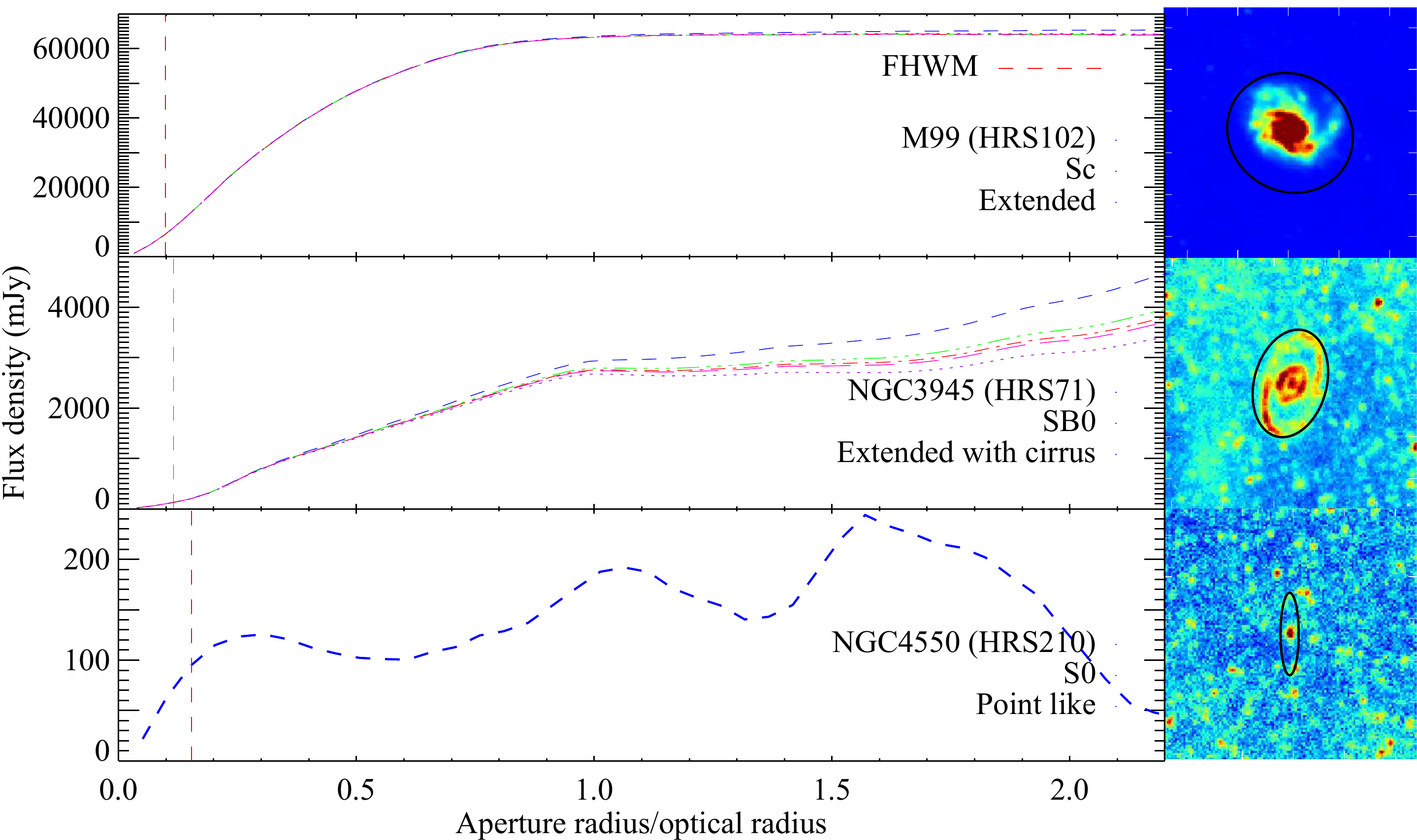} %
  		\caption{ \label{ouvdiff} Growth curves at 250~$\mu$m of the galaxies M~99 (HRS~102), NGC~3945 (HRS~71), and NGC~4550 (HRS~210). The different colored curves are obtained by changing the background estimate region as described in the text. The vertical red dashed lines correspond to the FWHM of the SPIRE beam at 250~$\mu$m. The right panels show the 250~$\mu$m images of the three galaxies, the black ellipses indicate the optical shapes of the galaxies.}
	\end{figure*}

	To understand and quantify the contribution of background features in the measurements, we choose three different objects as representative examples, as illustrated in Figure~\ref{ouvdiff}.
	There are three extreme cases.
	The first one is M99 (HRS~102) which is a bright resolved face-on spiral.
	The second one is NGC~3945 (HRS~71), a nearly face-on barred spiral, lying in a region polluted by a strong cirrus emission.
	The last one is NGC~4550 (HRS~210), an unresolved faint early-type galaxy.
	This source is treated as all point-like sources, but we choose to include it as a comparison with the two previous extended galaxies.
	Different growth curves are obtained when the background is estimated in different regions, as depicted in Figure~\ref{ouvdiff}.
	Here the different growth curves (colored lines) are obtained when the sky background is estimated in annuli of constant width of 60\arcsec\ but of 30\arcsec\ increasing radii.
	M99 (HRS~102) is a prototypical case with no particular problems since its growth curve reaches a plateau.
	Indeed, M99 is very bright, thus its flux density is not affected by faint background sources at large radii.
	On the contrary, the presence of Galactic cirrus strongly affects the flux density measurements, as for NGC~3945 (HRS~71).
	In this case, the growth curves do not saturate after a given radius.
	The curves of NGC~4550 (HRS~210) clearly show the contribution of background features at radii greater than $0.3\times a_{opt}$, where $a_{opt}$ is the optical semi major axis taken from NED.
	Furthermore, as NGC~4550 (HRS~210) is in a crowded field, several background sources contribute to any background region chosen.
	The HRS sample is composed of galaxies of different properties such as those shown in Figure~\ref{ouvdiff}.
	It is thus clear that a standard, automatic procedure cannot be blindly applied for the extraction of the flux densities of all the sources.
	We thus need to define appropriate apertures for each object.

	To define the apertures, we apply two different methods, one for the extended sources, mainly late-type galaxies, and another one for resolved but compact object, generally early-types.
	We inspect the infrared images and compare them with the optical ones.
	For most of the late-types, the infrared disk is more extended than its optical counterpart.
	We find that taking an elliptical diameter of 1.4 times the optical one is large enough to contain all the infrared emission of these galaxies.
	However, in some particular cases, this standard aperture needs to be adapted, especially for galaxies which are interacting, galaxies with a companion or a strong background source within the standard aperture.
	For instance, HI-deficient\footnote{The HI-deficiency is defined as the difference, in logarithmic units, between the HI mass expected from an isolated galaxy with the same morphological type and optical diameter and the observed HI mass \citep{HaynesGiovanelli84}.} spiral galaxies of the Virgo cluster have truncated dust disks \citep{Cortese10a}. 	
	Furthermore, edge-on spirals have an optical semi-minor axis $b_{opt}$ very small and  $1.4\times b_{opt}$ is not large enough to include the extended structure due to the side-lobes of the SPIRE beams (a typical example is NGC~4565-HRS~213 in Figure~\ref{allfigures}).
	We thus modify the ellipticity of edge-on galaxies of the aperture to include all the infrared emission.  
	
	Elliptical galaxies, even if resolved, have a faint compact infrared emission concentrated in the center.
	Lenticulars are the intermediate type between ellipticals and spirals.
	They contain extended dust but generally not beyond the optical emission.
	For all of these galaxies, the apertures are adapted to match the emission and avoid any major background contamination.

	For all galaxies, the background contribution is measured in a region defined as a circular annulus of inner radius $1.55\times a_{opt}$, with a 60\arcsec\ width.
	This choice is dictated by the fact that we want to quantify any possible contribution of large scale fluctuations in the sky background on source, but at the same time extract flux densities sufficiently far from the target to avoid any contamination of the galaxy to the sky background estimate.
	The width of the annulus is taken as a good compromise between the will of having a reliable statistic to estimate the background, and the will of avoiding as much as possible contamination from background sources and images features.
	We performed a detailed check on every background region to avoid or minimize any kind of source contamination (companion galaxy, strong background sources, etc).\\

 	The ellipses used for the aperture photometry and the circular annuli used for the background estimation are listed in Table~\ref{apertures}, organized as follows:
	\begin{itemize}
		\item Column 1: \textit{Herschel} Reference Survey name (HRS).
		\item Column 2: Zwicky name, from the Catalogue of Galaxies and of Cluster of Galaxies, \citep[][CGCG]{Zwicky61}. 
		\item Column 3: Virgo Cluster Catalogue name, \citep[][VCC]{Binggeli85}.
		\item Column 4: Uppsala General Catalogue name, \citep[][UGC]{Nilson73}.
		\item Column 5: New General Catalogue name, \citep[][NGC]{Dreyer1888}. 
		\item Column 6: Index Catalogue name, \citep[][IC]{Dreyer1895}. 
		\item Column 7: Right Ascension J2000 (RA). 
		\item Column 8: Declination J2000 (Dec).
		\item Column 9: Semi major axis of the aperture, in arcseconds ($a_{IR}$).
		\item Column 10: Semi minor axis of the aperture, in arcseconds ($b_{IR}$).
		\item Column 11: Position Angle, in degree (PA) (from North to East).
		\item Column 12: Inner radius of the background circular annulus, in arcseconds ($r_{in}^{bck}$).
		\item Column 13: Outer radius of the background circular annulus, in arcseconds ($r_{out}^{bck}$).
	\end{itemize}
	
	Figure~\ref{allfigures} shows the optical and infrared images of the HRS galaxies, along with the apertures used.
	Table~\ref{apratios} gives the mean and median infrared to optical aperture diameter ratios for elliptical, lenticular and late-type galaxies respectively, where the infrared diameter is the one listed in Table~\ref{apertures}.

	\begin{table}
	\centering
	\caption{ Median, mean and chosen infrared to optical aperture ratios of detected galaxies according to their morphological type. }
	\begin{tabular}{c c c c}
  	\hline\hline
  	  & Median ratio & Mean ratio & Chosen ratio \\
  	\hline 
  	E 					& 0.38 	& 0.29	& 0.30  \\ 
  	S0, S0a, S0/Sa 	& 0.76	& 0.88 	& 0.80\\ 
	Late-types 			& 1.40 	& 1.41 	& 1.40  \\ 
	\hline
	\label{apratios}
	\end{tabular}
	\tablefoot{The chosen ratios are used to calculate upper limits of sources not detected with SPIRE.}
	\end{table}

	\subsection{\label{aperture_corr}Aperture correction}
	
\begin{figure}
 	\includegraphics[width=\columnwidth]{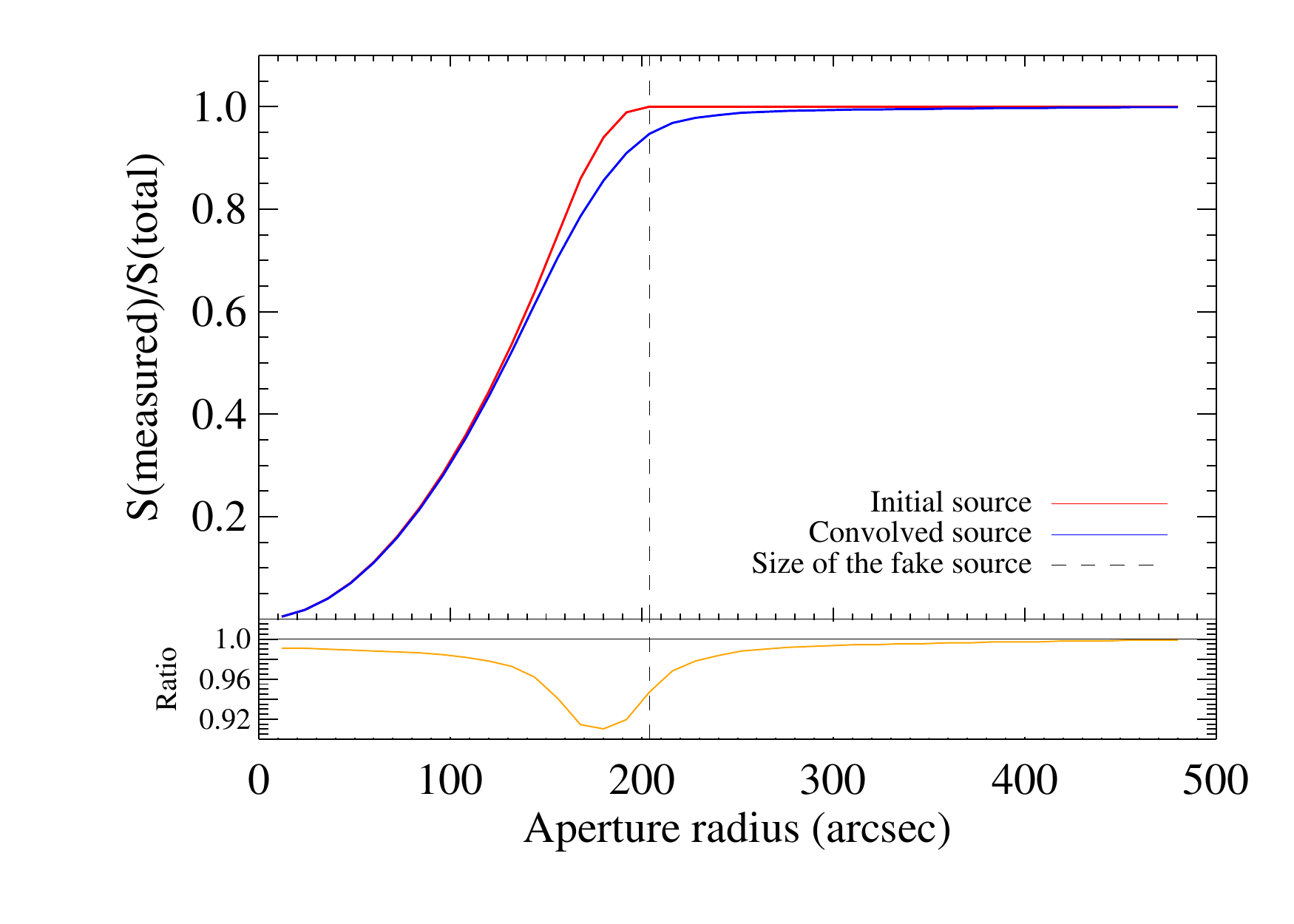}
 	\includegraphics[width=\columnwidth]{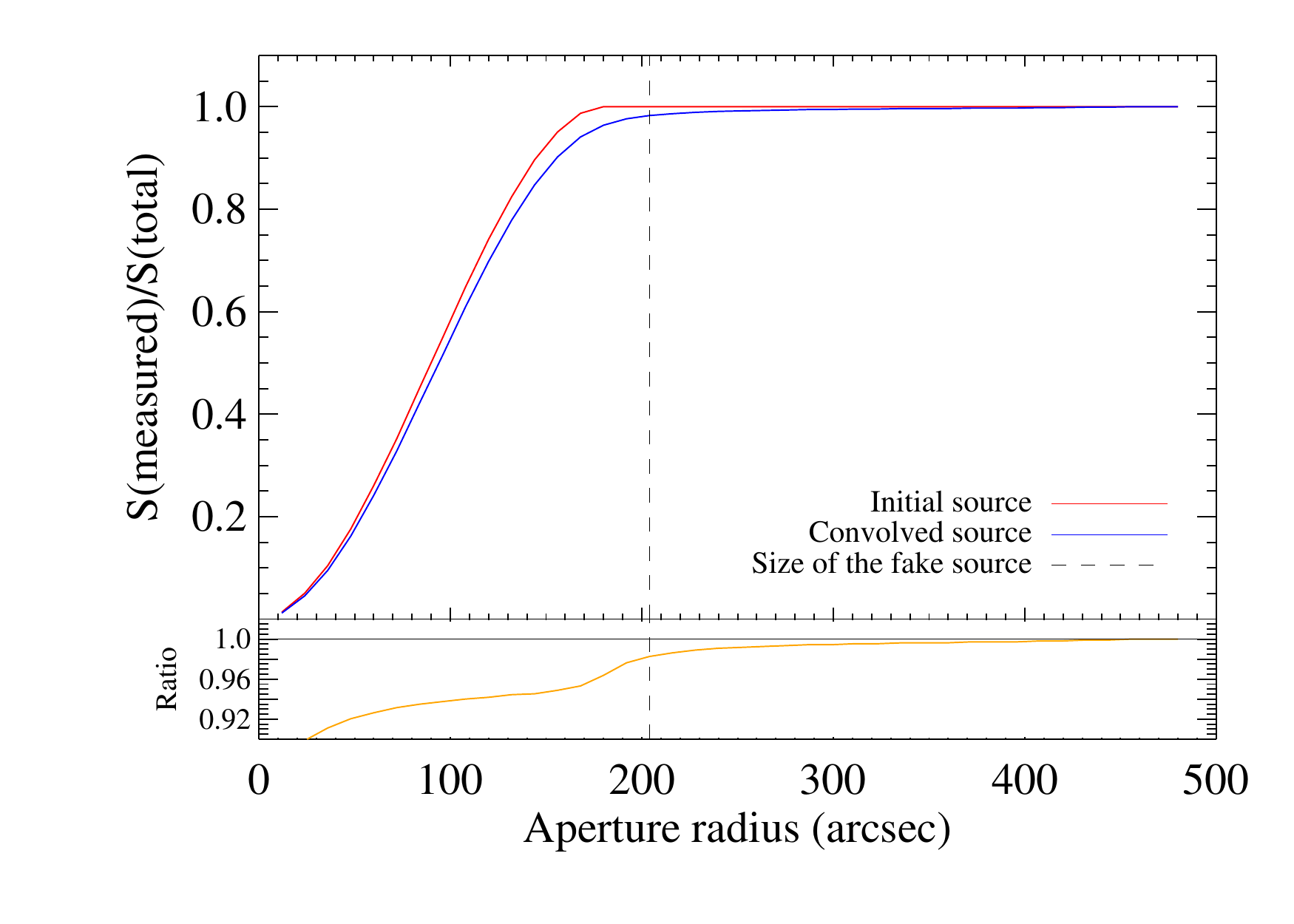}	
  	\caption{ \label{corr_ap} Simulation of the photometry of an extended face-on spiral galaxy at  500~$\mu$m. Upper panel: a flat extended source. In red, the integrated radial profile of the original source, and in blue of the convolved source. In orange, the convolved to original source flux density ratios. The dashed line marks the size of the original source. Lower panel: an extended galaxy with a linear surface brightness profile. }
 \end{figure}
 	
	As defined in Section~\ref{ext}, these apertures have been expressly chosen to include all of the infrared emission of the galaxies.
	Studying the emission of extended galaxies observed with \textit{Herschel}, \cite{Dale12} have shown that, at low surface brightness, the shape of the PSF can affect the emission at the edge of any object.
	They empirically defined the aperture correction as the ratio between the flux density measured on the IRAC 8.0~$\mu$m unsmoothed image, and the flux density measured on the same image smoothed to a \textit{Herschel} band PSF.
	They found a median value of 1.0 at all wavelengths, with maximum corrections between 7\% and 13\% for SPIRE.
	To quantify the effect of the wings of the PSF on our measurements, and understand whether a specific correction is required, we do the following exercise.
	The maximal effect is expected for an extended galaxy with a flat radial profile and a sharp edge at 500~$\mu$m.
	We create a mock galaxy on an image with the \textit{Herschel} 500~$\mu$m resolution, and with a constant surface brightness of 1 Jy beam$^{-1}$ dropping to 0 at a radius of 204\arcsec.
	Using the 500~$\mu$m PSF provided by Sibthorpe et al.\footnote{\url{ftp://ftp.sciops.esa.int/pub/hsc-calibration/SPIRE/PHOT/Beams/}}, we convolve the mock galaxy with the SPIRE 500~$\mu$m PSF.
	We carry out the photometry using circular apertures from $12\arcsec$ to $480\arcsec$ in steps of $12\arcsec$ on both the original and convolved images (Figure~\ref{corr_ap}).
	The largest aperture correction is $\sim5\%$ at the radius of the original source (204\arcsec).
	For a linear decreasing surface brightness profile, more physical but still extreme, the correction drops to $<2\%$.
	These 2$\%$-5$\%$ corrections can be considered as an upper limit for our data because: i) a flat radial profile is quite unphysical and ii) our apertures have been expressly chosen larger than the infrared size of the galaxies.
	As we always choose the aperture greater than the infrared emission of the galaxy (except for galaxies in particular configurations like NGC~4567-HRS~215 and NGC~4568-HRS~216), the aperture correction is thus much smaller than $<2\%$.
	As the calibration errors ($\sim$7\%) are greater than the aperture corrections, we choose not to apply them on our measurements and consider our flux densities as integrated values.\\

	\subsection{\label{photunc}Photometric uncertainties}
	
	There are two sources of uncertainty when carrying out photometry on SPIRE images, the systematic errors due to the absolute flux calibration and the stochastic errors related to the flux extraction technique.
	The calibration errors are (1) the uncertainty on the models used to determine the flux density of Neptune (5$\%$), (2) a 2$\%$ random uncertainty that is measured from the standard deviation in the ratio of the measured
Neptune flux density to the model Neptune flux density.
	The resulting calibration error is  7$\%$ in all bands (\citealt{Swinyard10}; SPIRE Observer's Manual).
	Technically, the errors should add together quadratically, but the SPIRE team decided to use 7$\%$ as a conservative upper limit on the flux calibration.
	As the methods used for point-like and extended source photometry are different, the stochastic error estimation is computed in different ways.	
	
	\subsubsection{\label{plerrors}Point-like sources}

		The uncertainty for point-like sources is calculated by performing tests in which artificial point sources with the same flux density as the target were added to the timeline data at random locations within a $0.3\deg$ box centered on each source.  
		The artificial sources were then fit with the timeline-based source fitter using the same settings as were applied to each target galaxy.  
		A hundred iterations of adding artificial sources to the fields around each galaxy were performed, and the standard deviation of the flux densities of the artificial sources was used as the uncertainty in the flux density measurement of the target galaxy.
		The highest value of the point-like source errors is 5 mJy for sources with a flux density less than 200 mJy.
	
	\subsubsection{Extended sources}
	
	For aperture photometry of extended sources, the stochastic total error, $err_{\rm tot}$, depends mainly on (1) the instrumental error, $err_{\rm inst}$, (2) the confusion error, $err_{\rm conf}$ and (3) the error on the determination of the sky background, $err_{\rm sky}$.	
	We calculate the errors on our flux density measurements according to the formula:\\ 

	\begin{equation}
	\label{erroreq}
		err_{\rm tot}=\sqrt{ err_{\rm inst}^2 + err_{\rm conf}^2 + err_{\rm sky}^2 },
	\end{equation}

		\subsubsection*{The instrumental error: $err_{\rm inst}$}
		
		The instrumental error is due to the noise of the instrument which depends on the number of scans crossing a pixel.
		Assuming it independent from pixel to pixel, the instrumental error is:
		
		\begin{equation}
		\label{inst}
		err_{\rm inst}= \sqrt{ \sum^{N_{\rm pix}}_{\rm i=\bf1} \sigma_{\rm inst,i}^2},
		\end{equation}
		\noindent where $N_{\rm pix}$ is the number of pixels within the aperture and  $ \sigma_{\rm inst}$ is the pixel per pixel uncertainty measured in the aperture on the error map provided by the pipeline.
		Mean values of $err_{\rm inst}$ are 1.2$\%$, 1.4$\%$ and 2.4$\%$ of the total flux density at 250, 350 and 500~$\mu$m, respectively.
		
		\subsubsection*{The confusion error: $err_{\rm conf}$}				
	
		The confusion error is due to the presence of background sources (i.e. faint point-like sources) within the aperture. 
		As the beam size is larger than the pixel size, this uncertainty is correlated between neighboring pixels.
		A point-like background source will then affect several pixels.
		The confusion error is:	

		\begin{equation}
		\label{conf}		
		err_{\rm conf}= \sigma_{\rm conf}^{\rm \lambda} \times \sqrt{\frac{N_{\rm pix} \times  {pixsize_{\rm \lambda}^2}}{{beam_{\rm \lambda}}}},
		\end{equation}
		\noindent where $\sigma_{\rm conf}^{\rm \lambda}$ is the confusion noise. Here, we assume the values estimated by \cite{Nguyen10}, i.e. $5.8$, $6.3$ and $6.8$  mJy/beam at 250, 350 and 500~$\mu$m respectively.
		The pixel size of the images $pixsize_{\rm \lambda}$ and the beam area $beam_{\rm \lambda}$ are given in Table~\ref{corr}.
		Mean values of $err_{\rm conf}$ are 4.2$\%$, 5.8$\%$ and 9.2$\%$ of the total flux density at 250, 350 and 500~$\mu$m, respectively.
		They are thus dominant with respect to the instrument noise.

		\subsubsection*{The background error: $err_{\rm sky}$}

		The uncertainty on the sky background comes from large scale structures not removed during the map-making procedure.
		These large scale structures, for instance, can be due to Galactic cirrus, as those evident in Figure~1 of \cite{Davies12} in the Virgo Cluster. 
		Indeed, despite the fact that the galaxies are selected at high Galactic latitude, some  images are contaminated by cirrus (see Figure~\ref{ouvdiff}).
		They contribute to the galaxy emission and/or to the background determination.
		To determine $ \sigma_{\rm sky}$, the uncertainty on the background, we take $13\times13$ pixel boxes around the galaxy in the image map for all of the three bands, we calculate the standard deviation of the mean values of the same boxes, as described in \cite{boselli03}.
		Ideally we would estimate $ \sigma_{\rm sky}$ from boxes with a similar number of pixels to the apertures used for the photometry. 
		This was not possible due to the sizes of the images. 
		The effect of using smaller boxes will be to give us a conservative estimate of $ \sigma_{\rm sky}$.
		The number of boxes depends on the size of the galaxy and on the size of the image; the mean numbers of boxes are 16, 14 and 11 at 250, 350 and 500~$\mu$m. 
		The error on the sky determination is:
		
		\begin{equation}
		\label{sky}
		err_{\rm sky}= N_{\rm pix} \sigma_{\rm sky},
		\end{equation}
		\noindent where $ \sigma_{\rm sky}$ is the uncertainty of the background.
		Mean values of $err_{\rm sky}$ are 8.0$\%$, 9.6$\%$ and 10.3$\%$ of the total flux density at 250, 350 and 500~$\mu$m, respectively.
		$err_{\rm sky}$ is thus the dominant error for extended galaxies. 
\newline

\begin{figure}
 		\includegraphics[width=\columnwidth]{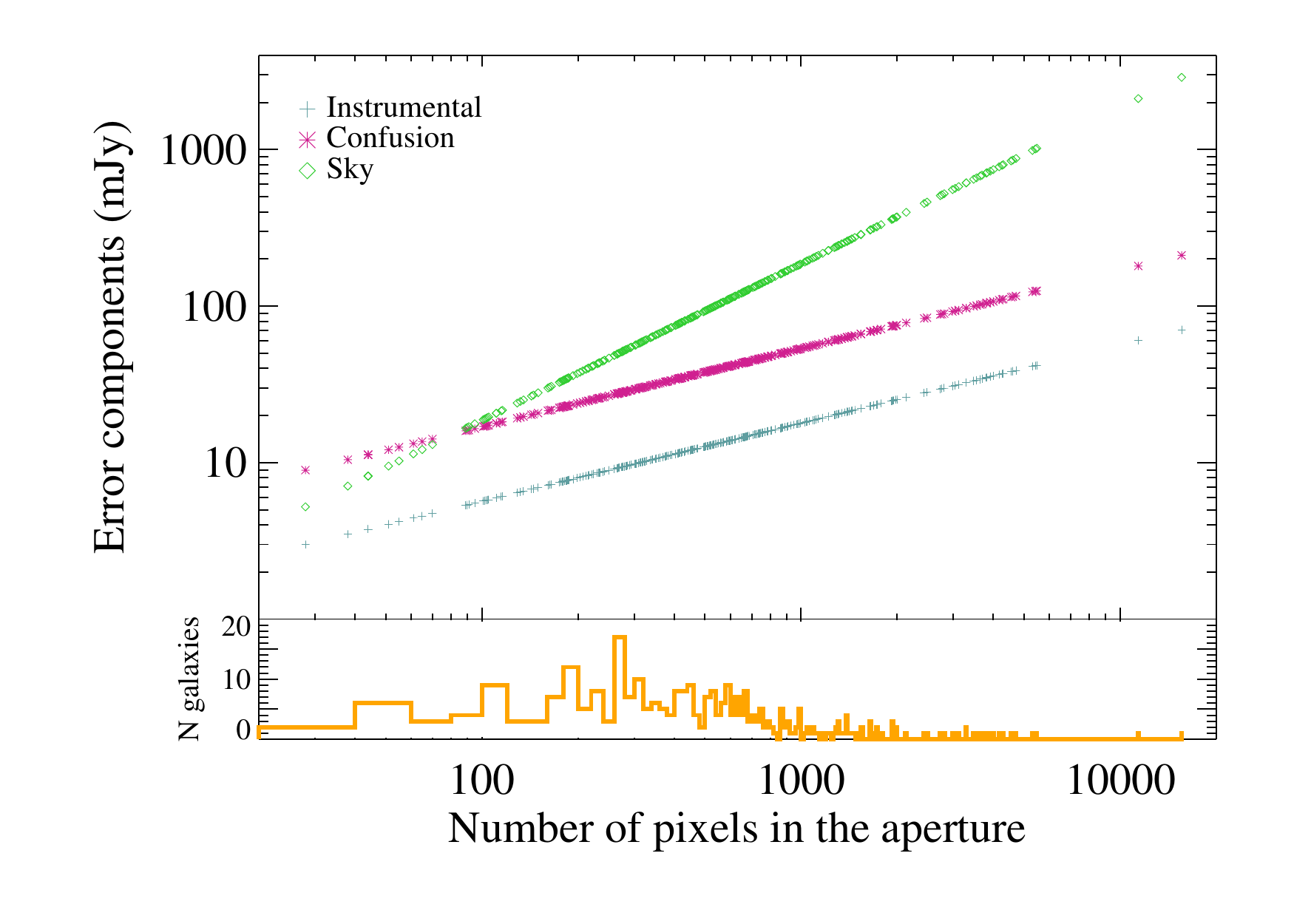}
  		\caption{ \label{errors} Error components versus the number of pixels of the aperture and the histogram of the number of pixels. }
	\end{figure}

	Figure~\ref{errors} shows the influence of each error component as a function of the number of pixels of the aperture, assuming mean values of $\sigma_{\rm inst}$ and $\sigma_{\rm sky}$ ($\sigma_{\rm conf}$ is constant at a given band).
	The background error is the dominant source of uncertainty for extended galaxies  of size larger than 80 pixels, which is the case for more than 90$\%$ of the galaxies of our sample.	
	
		\subsubsection*{Independent measurements}
	
\begin{figure}
 		\includegraphics[width=\columnwidth]{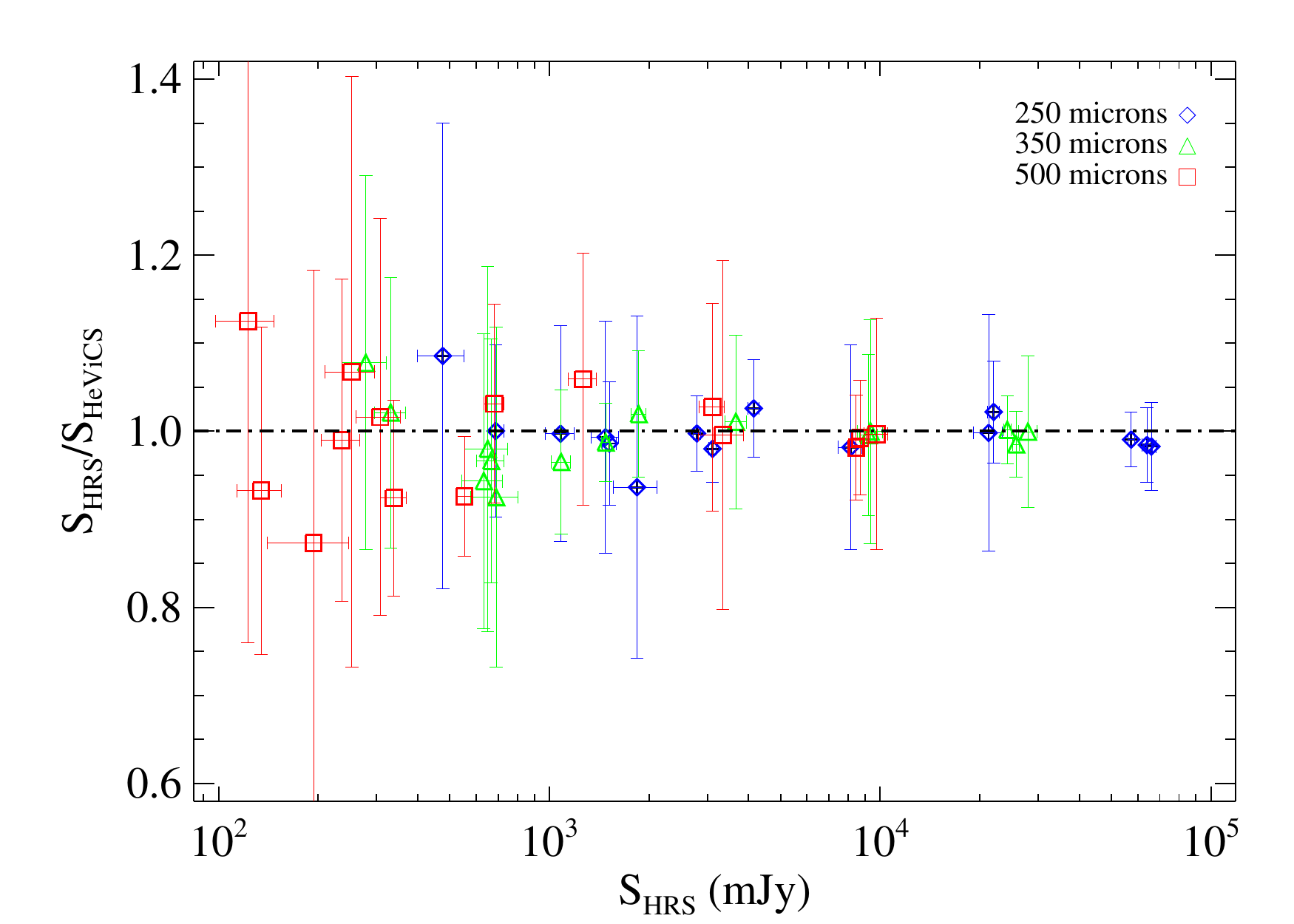}
  		\caption{ \label{hrshevics} Ratio between flux density measurements of 15 galaxies from HRS images and measurements of the same 15 galaxies from HeViCS images. Blue diamonds are for 250~$\mu$m flux densities, green triangles are for 350~$\mu$m and red squares are for 500~$\mu$m. }
	\end{figure}

	As part of the Science Demonstration Phase, 15 extended galaxies were observed in both the HRS and HeViCS projects.
	To test the reproducibility of our measurements and the accuracy of our errors, we compare the flux densities of the two sets of data.
	Indeed, we have two sets of independent images of the same galaxies produced by two different SPIRE scan modes.
	We perform the photometry on these two sets and compare the flux densities measured in exactly the same conditions (same apertures, same background regions and same photometric procedure).
	The names and flux densities of these sources from both HRS and HeViCS data are listed in Table~\ref{sdp}. 
	Figure~\ref{hrshevics} shows the ratio between the flux densities from HRS images and the flux densities from HeViCS images in the three bands.
	The median differences between the flux densities are 1.6$\%$, 1.9$\%$ and 3.0$\%$ at 250, 350 and 500~$\mu$m, respectively.
	We can consider these values as a lower limits to the photometric uncertainty on the flux densities measured in this work.\\
	
	\begin{table*}
	\centering
	\caption{The 15 extended galaxies observed in both HRS and HeViCS projects for the \textit{Herschel} Science Demonstration Phase. }
	\begin{tabular}{c c c c c c c c}
  	\hline\hline
  	  HRS 	& Name  &  \multicolumn{3}{c}{HRS} & \multicolumn{3}{c}{HeViCS}\\
  	   		&   		& 250~$\mu$m & 350~$\mu$m & 500~$\mu$m & 250~$\mu$m & 350~$\mu$m & 500~$\mu$m \\
  			& 		  & mJy & mJy & mJy & mJy & mJy & mJy \\
	\hline 
	  	102    & NGC~4254	&	$64026.1\pm  2329.3$	&$ 25753.9\pm  599.1$		&$8685.7\pm  406.2$		&$65039.2\pm  1525.8$	&$26143.0\pm  778.7$	&$8750.1\pm  399.3$ \\
  		106    & NGC~4276	&	$1476.4\pm  139.5$ 		& $650.7\pm  95.8$			&$252.7\pm  42.6$		&$1486.6\pm  137.9$		&$664.0\pm  100.9$		&$236.7\pm  62.7$	\\
  		122    & NGC~4321	&	$66006.4\pm  2209.8$	& $27948.0\pm  1792.0$		&$9742.4\pm  817.0$		&$67163.5\pm  2583.7$	&$27962.1\pm  1602.6$	&$9773.1\pm  995.4$	\\
  		152    &	NGC~4412	&	$2790.7\pm  73.6	$		& $1084.2\pm  71.7 $			&$338.6\pm  30.7$		&$2797.8\pm  95.1$		&$1123.4\pm  60.4$		&$366.4\pm  29.0$	\\
  		158    &	NGC~4423  	&	$1079.9\pm  107.3$		& $632.9\pm  88.8$			&$307.6\pm  47.4$		&$1082.9\pm  78.3$		&$670.9\pm  72.9$		&$302.7\pm  48.4$	\\
  		160    &	NGC~4430   &	$4145.2\pm  156.3$		& $1859.9\pm  96.1$			&$680.6\pm  45.8$		&$4040.2\pm  154.0$		&$1824.5\pm  87.7$		&$659.9\pm  57.2$	\\
  		162    &	NGC~4435   &	$1839.0\pm  276.7$		& $690.0\pm  112.8 $			&$193.5\pm  53.3$ 		&$1963.9\pm  280.9$		&$745.9\pm  96.9$		&$221.6\pm  49.7$	\\
  		163    &NGC~4438 	&	$8132.3\pm  684.6$		& $3660.7\pm   271.1$		&$1263.9\pm  122.7$		&$8284.0\pm  689.8$		&$3622.2\pm  230.6$		&$1193.3\pm  112.6$	\\
  		165    &UGC~7579  	&	$688.5\pm  38.8$			& $331.0\pm  35.5$			&$134.3\pm  20.7$		&$688.3\pm  55.0$ 		&$324.2\pm  34.2$		&$144.0\pm  18.2$	\\
  		182    &NGC~4480 	&	$3112.1\pm  84.2	$		& $1478.6\pm  42.8$			&$553.1\pm  31.7$		&$3175.4\pm  90.4$		&$1497.8\pm  52.2$		&$597.2\pm  27.3$	\\
  		190    &	NGC~4501	&	$57336.0\pm  1379.6$	& $24221.3\pm  645.1$		&$8461.8\pm  366.1$		&$57884.4\pm  1172.9$	&$24183.9\pm  666.2$	&$8623.7\pm  367.5$	\\
  		206    &	IC~3521	  	&	$1516.8\pm  72.8$		& $666.2\pm  63.3$			&$235.3\pm  31.3$		&$1538.2\pm  80.7$		&$689.5\pm  73.8$		&$237.7\pm  30.5$	\\
  		217    &	NGC~4569 	&	$22023.9\pm  894.2$		& $9219.8\pm  580.8$		&$3105.1\pm  265.6$		&$21550.7\pm  850.8$	&$9261.9\pm  619.2$		&$3022.1\pm  230.6$	\\
  		220    & NGC~4579  	&	$21263.2\pm   2091.8$ 	& $9340.7\pm  802.9$ 		&$3339.0\pm   526.2$	&$21301.7\pm  1950.3$	&$9343.3\pm  874.4$		&$3353.2\pm  405.0$	\\
  		223    & UGC~7802  	&	$475.8\pm  76.5$ 		& $278.6\pm  42.4$			&$122.5\pm  24.6$		&$438.3\pm  80.3$		&$258.5\pm  32.4$		&$108.9\pm  27.7$	\\
	\hline
	\label{sdp}
	\end{tabular}
	\end{table*}

	Table~\ref{err} indicates the mean total stochastic error for the extended sources within the HRS, for early-types and late-types galaxies separately, and for different flux density ranges.
	Calibration errors are not included.
	Adding the calibration errors, the mean total errors are 9, 11 and 13$\%$ at 250, 350 and 500~$\mu$m, respectively, which is consistent with the 10, 10 and 15$\%$ estimated by \cite{Davies12}. \\	
		
	\begin{table}
	\centering
	\caption{Mean stochastic errors ($err_{\rm tot}$) on the HRS flux densities for extended galaxies. }
	\begin{tabular}{c c c c}
  	\hline\hline
  	  & 250~$\mu$m  & 350~$\mu$m & 500~$\mu$m\\
  	\hline 
  	All 							& 6.2$\%$		& 8.2$\%$		& 11.1$\%$  \\ 
	\hline
  	E  							& 20.2$\%$		& 25.8$\%$ 	& 20.9$\%$ \\
	S0, S0a, S0/Sa				& 9.0$\%$		& 13.2$\%$		& 19.1$\%$	 \\	
	Late-types 					& 5.9$\%$		& 7.6$\%$		& 10.5$\%$ \\ 
	\hline
	$S<200$ mJy					& 21.8$\%$		& 24.6$\%$		& 21.6$\%$			\\
	$200$ mJy $< S<1000 $mJy	& 10.2$\%$		& 10.9$\%$		& 10.2$\%$			\\
	$S>1000$ mJy				& 4.6$\%$			& 5.9$\%$			& 5.9$\%$ 			\\
	\hline
	\label{err}
	\end{tabular}
	\end{table}

	\subsection{\label{und}Undetected sources}
	
	Bona fide detected galaxies are identified through a visual inspection of the images rather than following strict signal to noise criteria. 
	This choice is dictated by the fact that, given the different nature of the sky background and of the emitting source, we might have strong detections but with
a very low signal to noise (this is for instance the case of HRS~71 which has an uncertain flux density measurement since lying in a cirrus dominated region) or very high signal to noise sources with uncertain values (point like sources which can be easily confused with background objects). 
	If we limit our sample to extended sources with no cirrus contamination nor nearby companions, our detection threshold is S/N $\sim$ 3, 2 and 2 at 250, 350 and 500~$\mu$m, respectively, where the S/N is defined as the ratio of the flux density $S$ over the total uncertainty $err_{tot}$.
	
	For the undetected galaxies (39, 42 and 47 galaxies at 250, 350 and 500~$\mu$m, respectively), an upper limit is determined as:
	
	\begin{equation}
	S_{\rm limit(\lambda)}=3 err_{\rm tot},
	\end{equation}
	\noindent where $err_{\rm tot}$ is estimated as in Equation~\ref{erroreq}.
	The measure of $err_{\rm tot}$ requires the adoption of a representative aperture, $N_{\rm pix}$, for each undetected source.
	We make three different assumptions according to the morphology of the undetected galaxies.
	We form 3 groups: (a) type E, (b) type S0, S0a, S0/Sa and (c) late-types.
	For both group (a) and (b), we calculate the ratio between the semi-major axis of the infrared elliptical aperture and the semi-major axis of the optical diameter of the detected galaxies of the same morphological type (Table~\ref{apratios}).
	Given the mean values measured for detected sources, we decided to adopt $0.3\times a_{opt}$ for ellipticals and $0.8\times a_{opt}$ for S0 and S0/Sa.
	We take $1.4\times a_{opt}$ for late-type galaxies, as we do for detected late-types.
	The radii of the circular region used for the calculation of upper limits are then 0.3, 0.8  or 1.4 times the optical semi-major axis for ellipticals, lenticular and spiral galaxies, respectively.
	For galaxies detected only in one or two bands, but not in the others, we use the aperture defined at these bands and take $a_{IR}$ as the radius of the circular aperture to calculate the upper limit in the other bands.
	A minimum conservative and independent of $\lambda$ radius for upper limit apertures has been set to $22.5\arcsec$ not to have apertures smaller than the SPIRE resolution.

	\begin{table}
	\centering
	\caption{Detection rates in each band for different morphology classes.}
	\begin{tabular}{c c c c}
  	\hline\hline
  	  & 250~$\mu$m  & 350~$\mu$m & 500~$\mu$m\\
  	\hline 
  	E 					& 32$\%$	& 32$\%$	& 23$\%$  \\ 
  	S0, S0a, S0/Sa 	& 60$\%$	& 58$\%$ 	& 55$\%$ \\ 
	Late-types 			& 97$\%$	& 96$\%$	& 95$\%$  \\ 
	\hline
	\label{detectrate}
	\end{tabular}
	\end{table}
	
	\begin{figure}
 	\includegraphics[width=9cm]{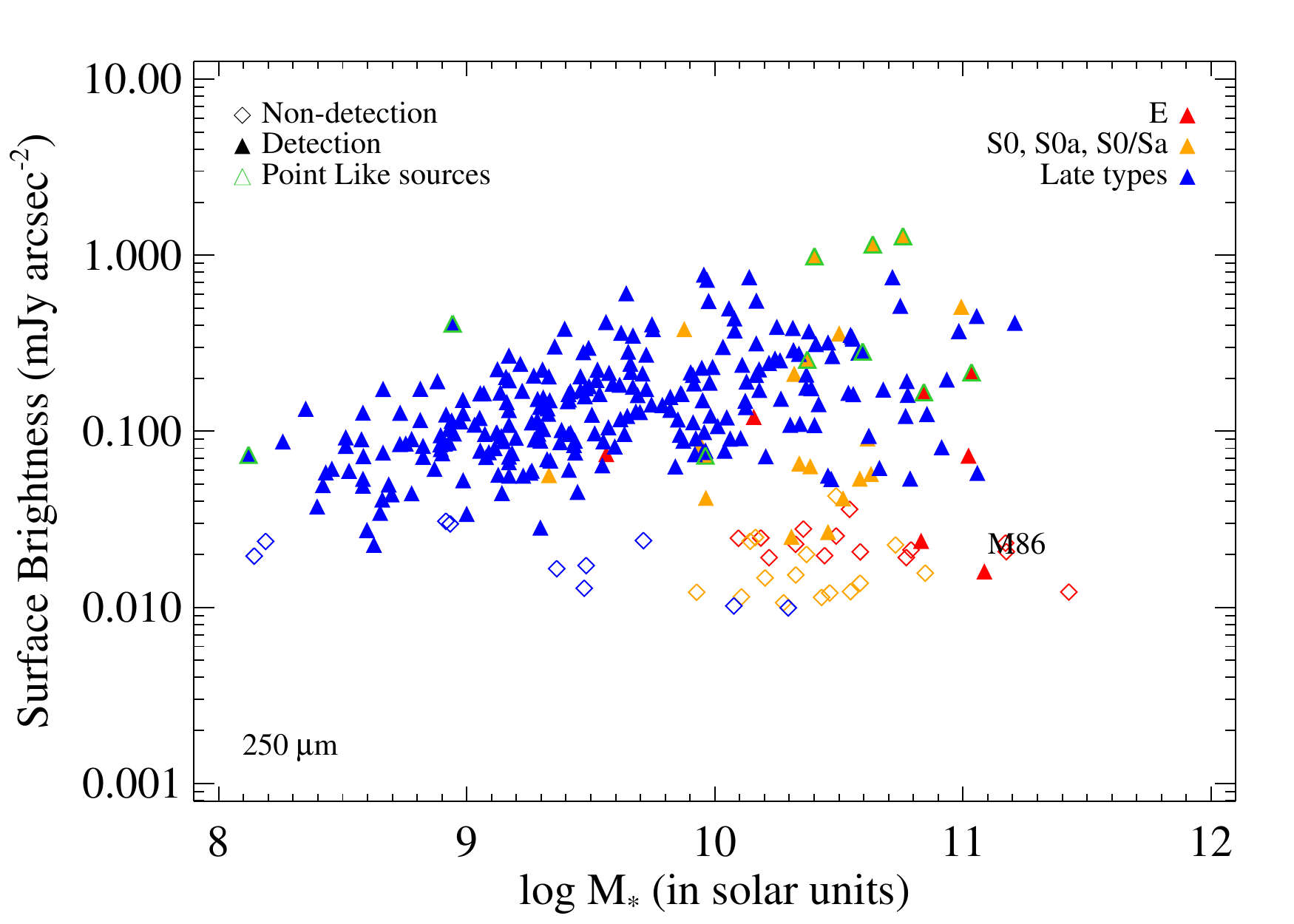}
 	\includegraphics[width=9cm]{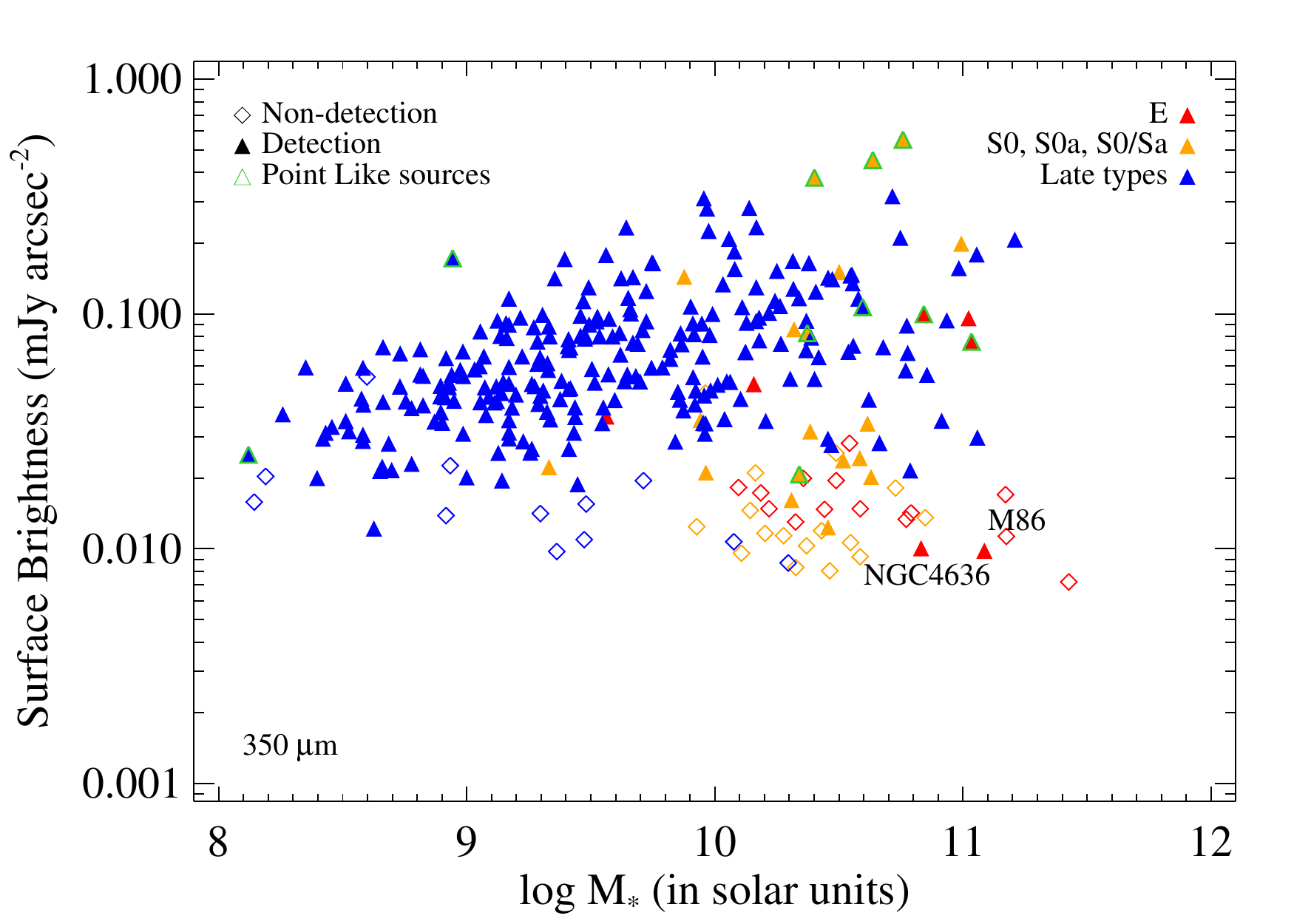}
	\includegraphics[width=9cm]{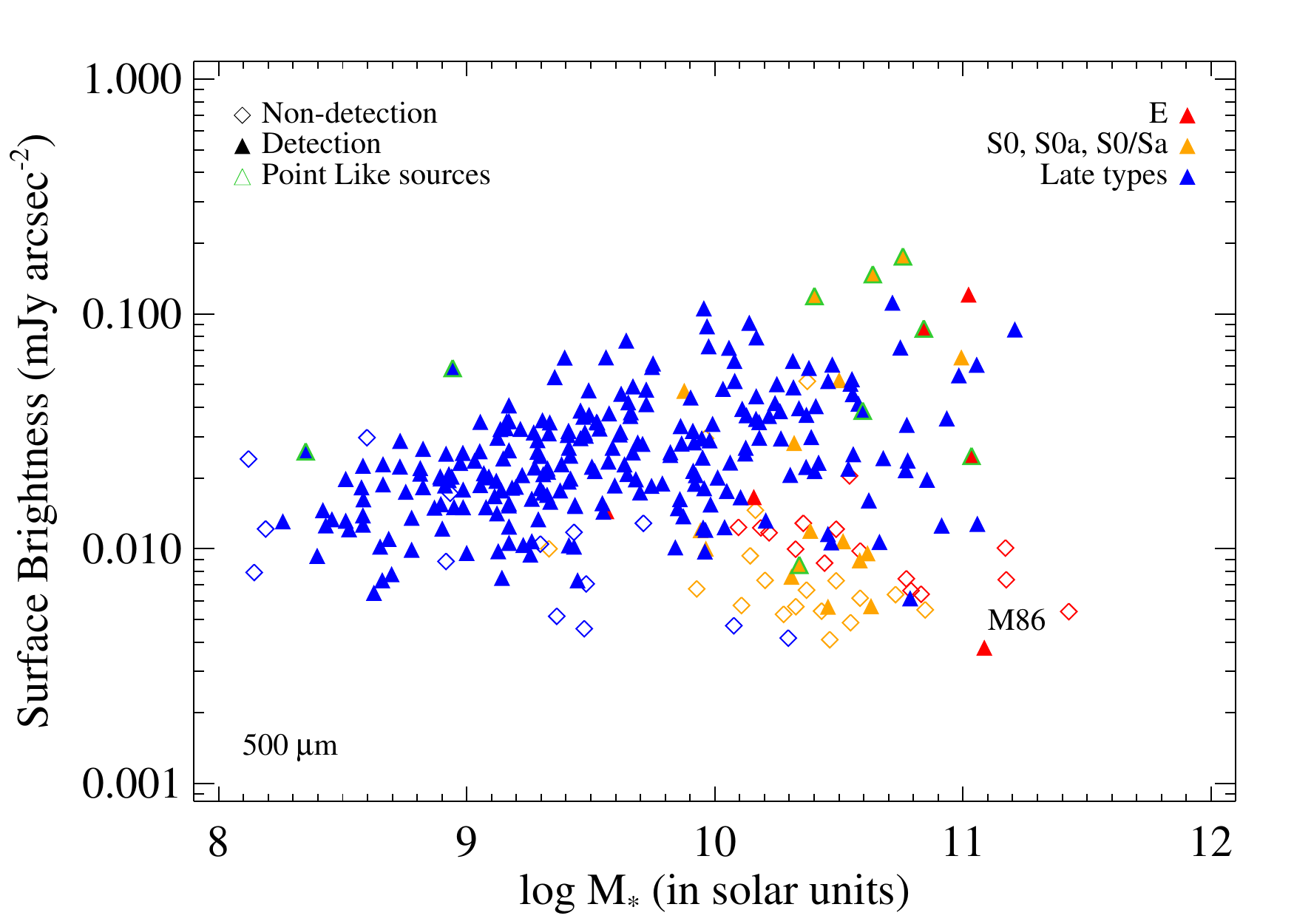}
  	\caption{ \label{surfhlum} The surface brightness versus the stellar mass at 250 (top panel), 350 (middle panel) and 500~$\mu$m (bottom panel). Filled triangles are for detections, empty triangles for non-detections. Red, orange and blue colors are for elliptical, lenticular and late-type galaxies, respectively. Triangles with a green contour are point-like sources. }
 	\end{figure}
	
	To test whether these upper limits are realistic, we plot in Figure~\ref{surfhlum} the surface brightness of the galaxies versus their stellar mass calculated as in \cite{Boselli09} \citep[for details, see][]{Boselli12}. 
	The surface brightness is calculated by dividing flux densities by the infrared size of galaxies, i.e. the aperture size.
	The detection limit in surface brightness of our survey is $\sim$ 0.03, 0.02 and 0.008 mJy arcsec$^{-2}$ at 250, 350 and 500~$\mu$m, respectively. 
	The only extended galaxy with a surface brightness below this threshold is M86 (HRS150) whose dust emission comes from a peculiar feature probably stripped from a nearby star forming system \citep{Gomez10}.
	At 350~$\mu$m, the other detected galaxy with a low surface brightness is NGC4636 (HRS241) which is a faint, clearly detected, compact source.
	Few sources have a relatively high surface brightness at 250~$\mu$m but are non detected at 350 and 500~$\mu$m.
	These sources are close to point-like and are well detected at 250~$\mu$m.
	However, at 350 and 500~$\mu$m, they become as faint as the background sources, with a comparable surface brightness.
	To be conservative, we thus consider them as undetected sources.
	These two galaxies are particular cases, thus we consider that our upper limits are realistic as they lie at the lower limit of the detections.\\
	
	Table~\ref{detectrate} gives the detection rate in each band for the 3 groups: ellipticals (E), lenticulars (S0, S0a, S0/Sa) and late types.

\section{\label{fluxdens}Flux densities calculation}

SPIRE maps are in Jy/beam. Flux densities, upper limits and errors of extended sources are thus converted into mJy using Equation~\ref{conv}.

\begin{equation}
\label{conv}
S_{\rm \lambda}(mJy)= S_{\rm \lambda}(Jy/beam) \frac{pixsize_{\rm \lambda}^2 \times 1000 }{beam_{\rm \lambda}} \times corr_{\rm \lambda},
\end{equation}
\noindent where the pixel size of the images, $pixsize_{\rm \lambda}$, the beam area, $beam_{\rm \lambda}$, and the correction, $corr_{\rm \lambda}$ (latest HIPE v8 calibration\footnote{\url{http://herschel.esac.esa.int/twiki/bin/genpdf/Public/HipeWhatsNew8x?pdforientation=portrait&pdftoclevels=3}} and extended sources corrections) are given in Table~\ref{corr}.

	\subsection{The data table}

The flux densities of the 323 HRS galaxies in the three SPIRE bands (without colour corrections applied) are given in Table~\ref{HRS_phot}, organized as follows:

\begin{itemize}
	\item Column 1: HRS name.
	\item Column 2: Flag of the 250~$\mu$m  flux density ($f_{250}$); 0: Non-detection, 1: Detection (extended source), 2: Detection (point-like source),  3: Overestimation of the flux density due to the presence of a strong background source or a companion galaxy, 4: Presence of Galactic cirrus.
	\item Column 3: Flux density at 250~$\mu$m ($S_{250}$) in mJy.
	\item Column 4: Flag of the 350~$\mu$m  flux density ($f_{350}$); see Column 2.
	\item Column 5: Flux density at 350~$\mu$m ( $S_{350}$) in mJy.
	\item Column 6: Flag of the 500~$\mu$m  flux density ($f_{500}$); see Column 2. 
	\item Column 7: Flux density at 500~$\mu$m ($S_{500}$) in mJy. 
	\item Column 8: Number of pixels in the 250~$\mu$m  aperture ($N_{250}$).
	\item Column 9: Number of pixels in the 350~$\mu$m  aperture ($N_{350}$). 
	\item Column 10: Number of pixels in the 500~$\mu$m  aperture ($N_{500}$). 
	\item Column 11: Instrumental error at 250~$\mu$m, determined as in Equation~\ref{inst} ($err^{\rm 250}_{\rm inst}$).
	\item Column 12: Instrumental error at 350~$\mu$m, determined as in Equation~\ref{inst} ($err^{\rm 350}_{\rm inst}$). 
	\item Column 13: Instrumental error at 500~$\mu$m, determined as in Equation~\ref{inst} ($err^{\rm 500}_{\rm inst}$). 
	\item Column 14: Confusion error at 250~$\mu$m, determined as in Equation~\ref{conf} ($err^{\rm 250}_{\rm conf}$).
	\item Column 15: Confusion error at 350~$\mu$m, determined as in Equation~\ref{conf} ($err^{\rm 350}_{\rm conf}$). 
	\item Column 16: Confusion error at 500~$\mu$m, determined as in Equation~\ref{conf} ($err^{\rm 500}_{\rm conf}$). 
	\item Column 17: Sky error at 250~$\mu$m, determined as in Equation~\ref{sky} ($err^{\rm 250}_{\rm sky}$).
	\item Column 18: Sky error at 350~$\mu$m, determined as in Equation~\ref{sky} ($err^{\rm 350}_{\rm sky}$). 
	\item Column 19: Sky error at 500~$\mu$m, determined as in Equation~\ref{sky} ($err^{\rm 500}_{\rm sky}$). 														
\end{itemize}	

The  $err_{\rm inst}$,  $err_{\rm conf}$ and $err_{\rm sky}$ are not provided for point-like galaxies (i.e. flag 2) as their errors are calculated with a different method (see Section~\ref{plerrors}).
The total errors provided in Table~\ref{HRS_phot} do not contain the calibration error of 7$\%$ which can be added in quadrature.
\section{\label{comp}Comparison with the literature}

	Submillimetre photometry in the $\sim$ 250-550~$\mu$m spectral domain is available for some HRS galaxies from \cite{Davies12}, Auld et al. (submitted), \cite{Dale12} and \cite{Planckcatalogue}.
	 
	\subsection*{Comparison with the HeViCS Bright Galaxy Catalogue}
\begin{figure}
 	\includegraphics[width=9cm]{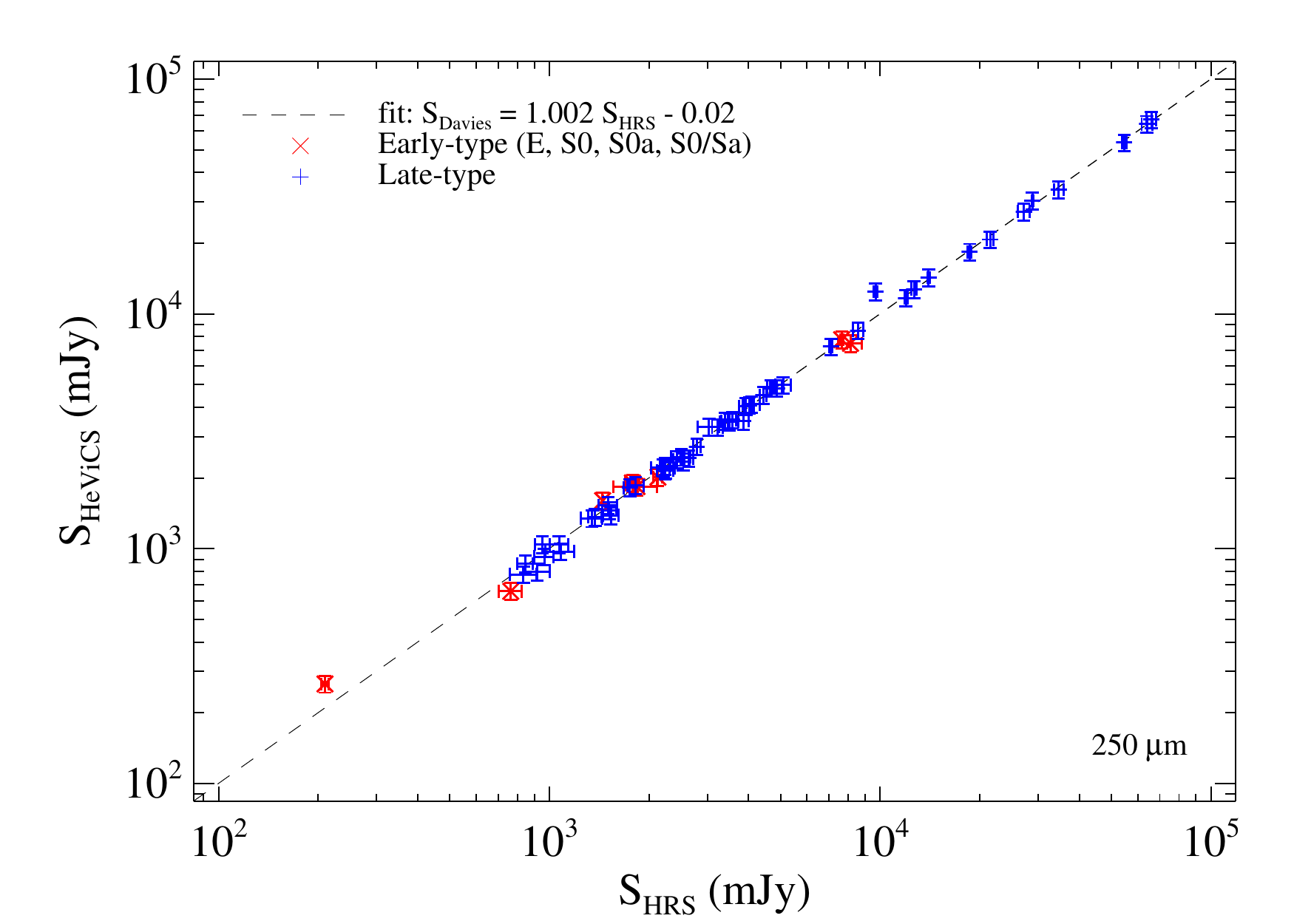}
 	\includegraphics[width=9cm]{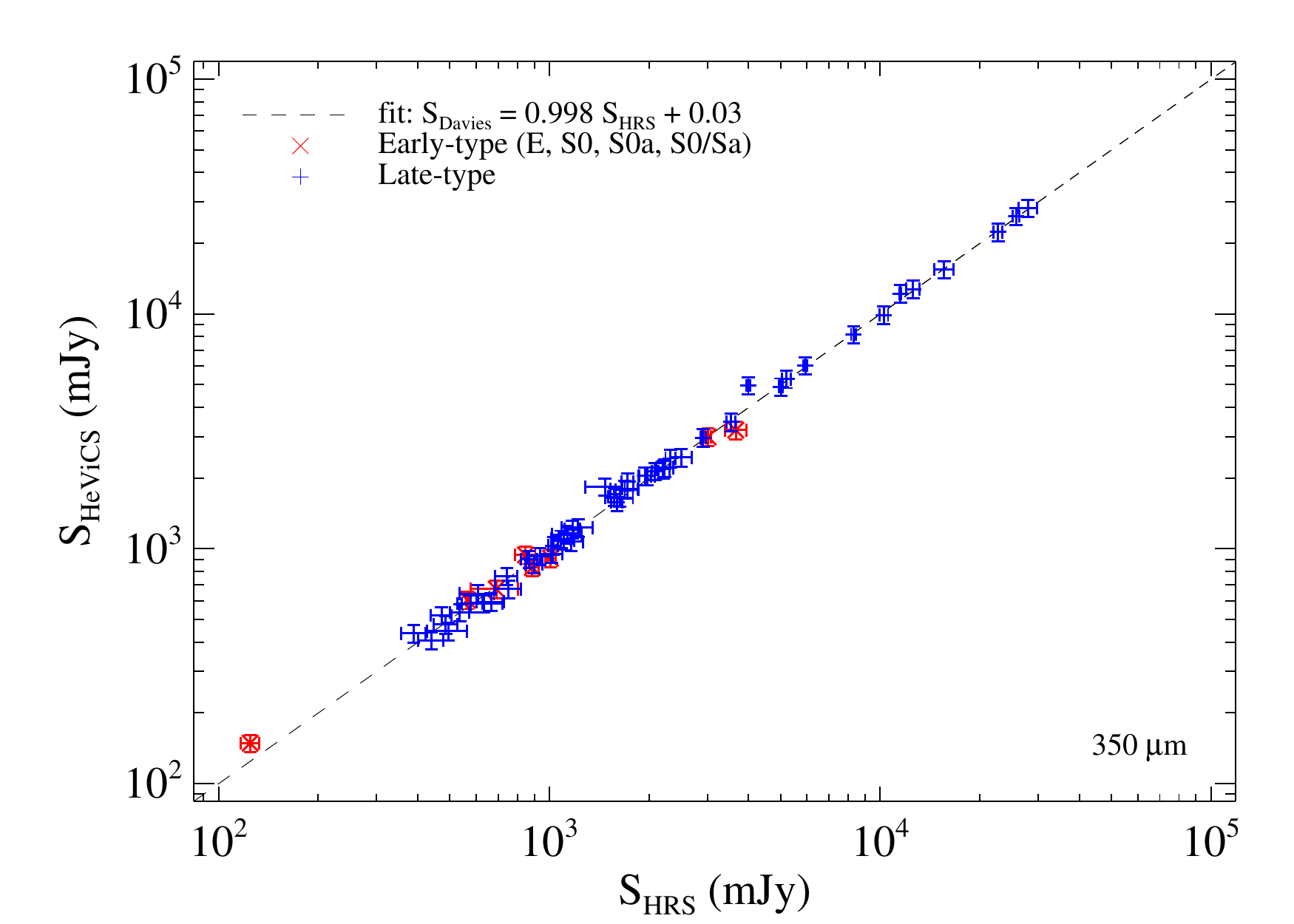}
	\includegraphics[width=9cm]{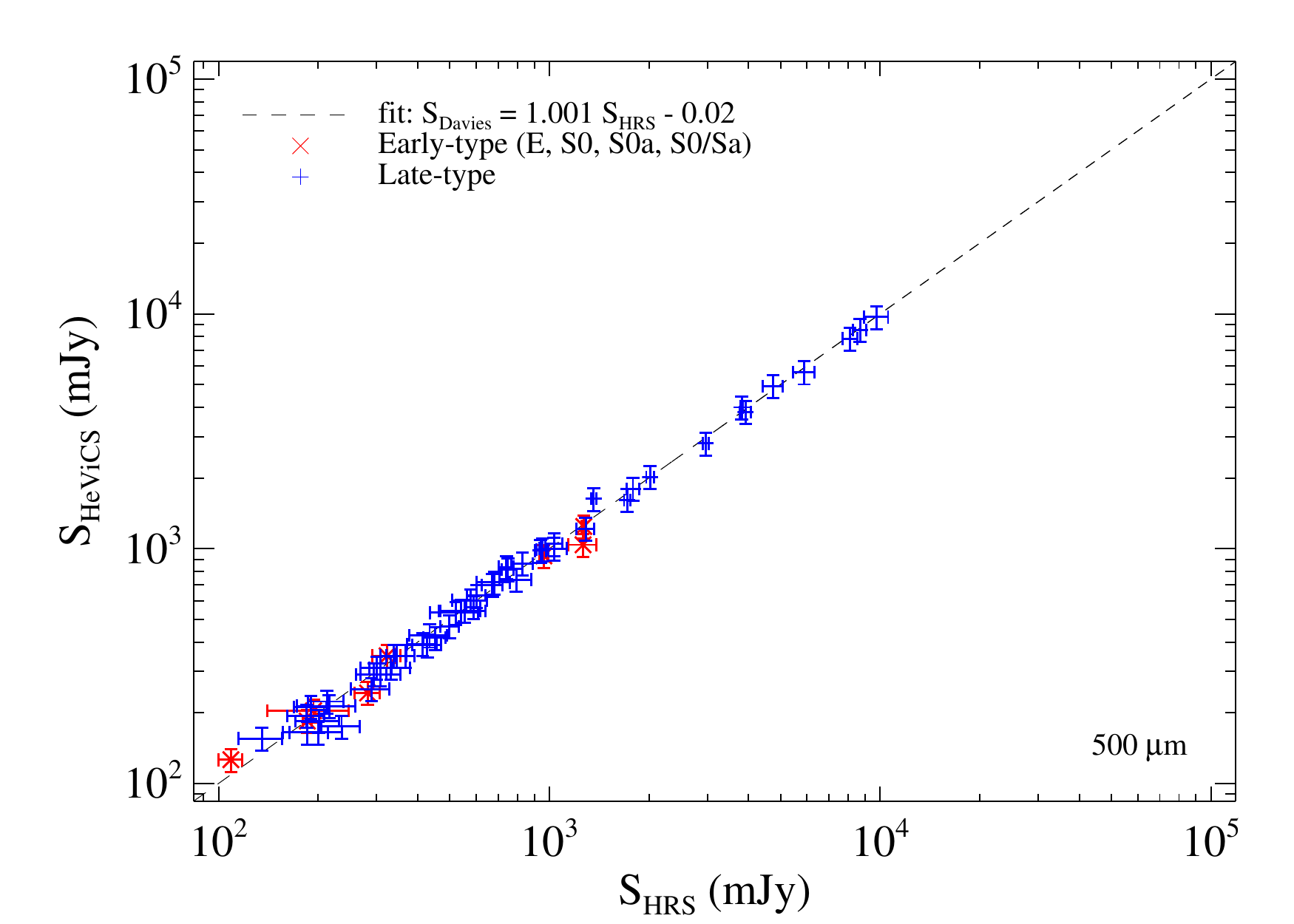}
  	\caption{ \label{daviesvsme} A comparison of the flux densities of the 59 sources common to both the HRS sample and the HeViCS Virgo bright galaxy sample \citep{Davies12} at 250, 350 and 500~$\mu$m. Red triangles are for early type galaxies, blue triangles for late type galaxies. The dashed line indicates the linear fit. }
\end{figure}

We compare our results with those of \cite{Davies12} (Figure~\ref{daviesvsme}) who used a different method to perform the photometry. 
They carried out a study on 78 bright Virgo galaxies as part of the HeViCS.
Before extracting flux densities, they smoothed and re-gridded the 250~$\mu$m and 350~$\mu$m images to the 500~$\mu$m resolution and pixel scale.
They defined ``by eye'' elliptical apertures and a concentric annulus (for background estimation) on the 500~$\mu$m image and used them at all bands.
For consistency with our work, we apply the $K_4$ corrections to their measurements, using the values given in Table~\ref{corr}.
For the 59 galaxies in common to both the HRS and HeViCS surveys, mean values of the flux density ratios between their measurements on HeViCS fields and in this work are $1.00 \pm 0.07$, $1.01 \pm 0.07$ and $0.99 \pm 0.08$ at 250, 350 and 500~$\mu$m.
Thus our fluxes and those of \cite{Davies12} are consistent with each other.
Auld et al. (2012, submitted) present a comparison between the flux densities of the optically selected Virgo galaxies, from the Virgo Cluster Catalogue (VCC), and HRS ones, for galaxies in common.
Despite the different techniques used (automatic for Auld et al. 2012), the measurements are in good agreement.
\\

	\subsection*{Comparison with KINGFISH}
	
	KINGFISH (Key Insights on Nearby Galaxies: A Far-Infrared Survey with \textit{Herschel}; \citealt{Kennicutt12}) is a survey of 61 nearby galaxies observed in PACS and SPIRE bands.
	\cite{Dale12} provide the flux densities of this sample in which six galaxies are in common with the HRS.
	The comparison is important as the images of the targets are the same but the data reduction, map-making, and flux extraction use different methods.
	They carried out aperture photometry using ellipses and applied aperture correction, which are of the order of a few percents (Dale, private communication). 
	They also applied Galactic extinction corrections to their measurements. 
	These corrections are however very small since they do not exceed 0.4$\%$, where this value has been determined for an object $b=10\deg$.
	The background is estimated by taking the mean value of several regions circumscribing the galaxy.
	The $K_4$ correction is applied to their flux densities to have consistency between their and our measurements.
	
\begin{figure*}
 	\includegraphics[width=\columnwidth]{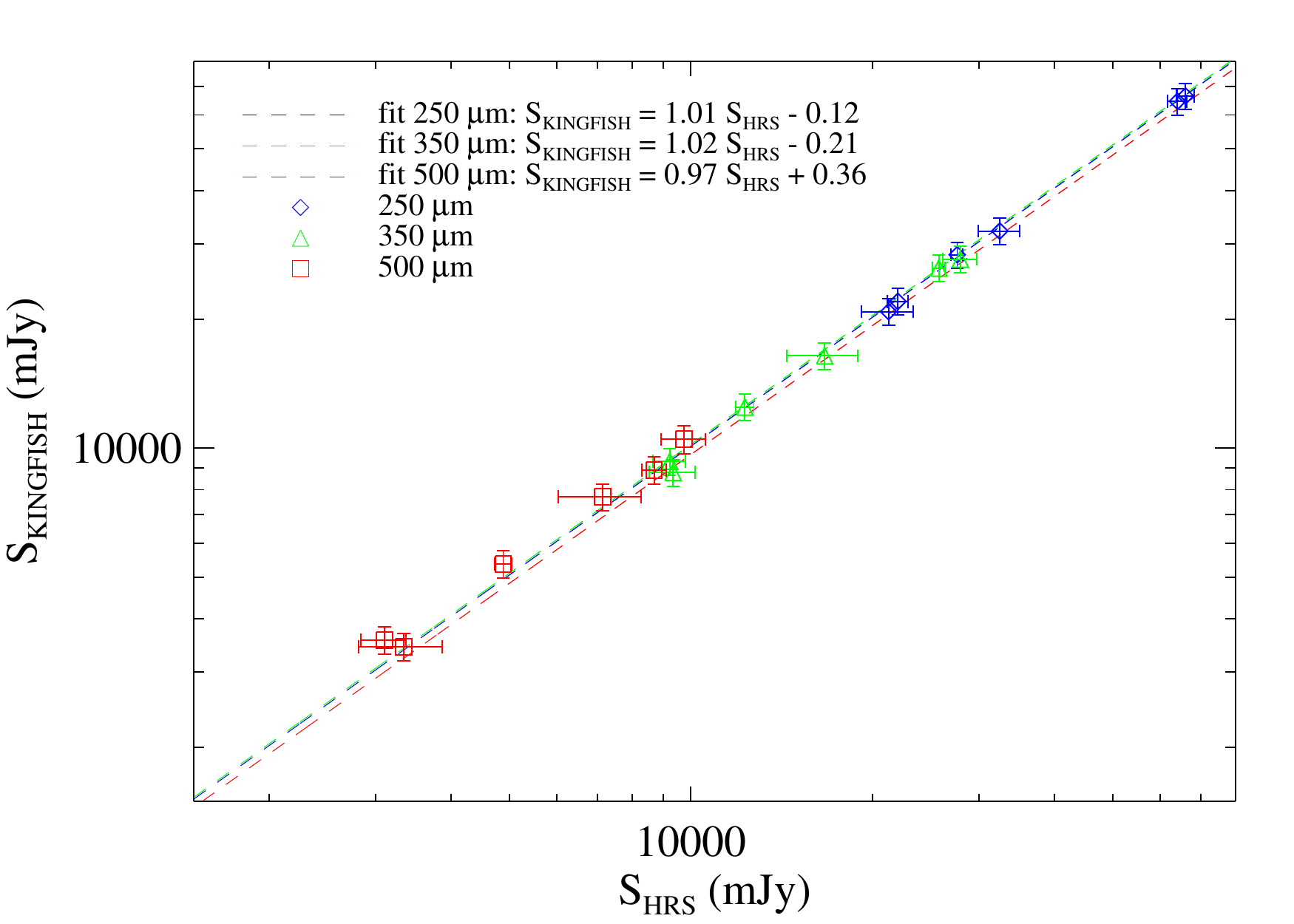}\hfill
 	\includegraphics[width=\columnwidth]{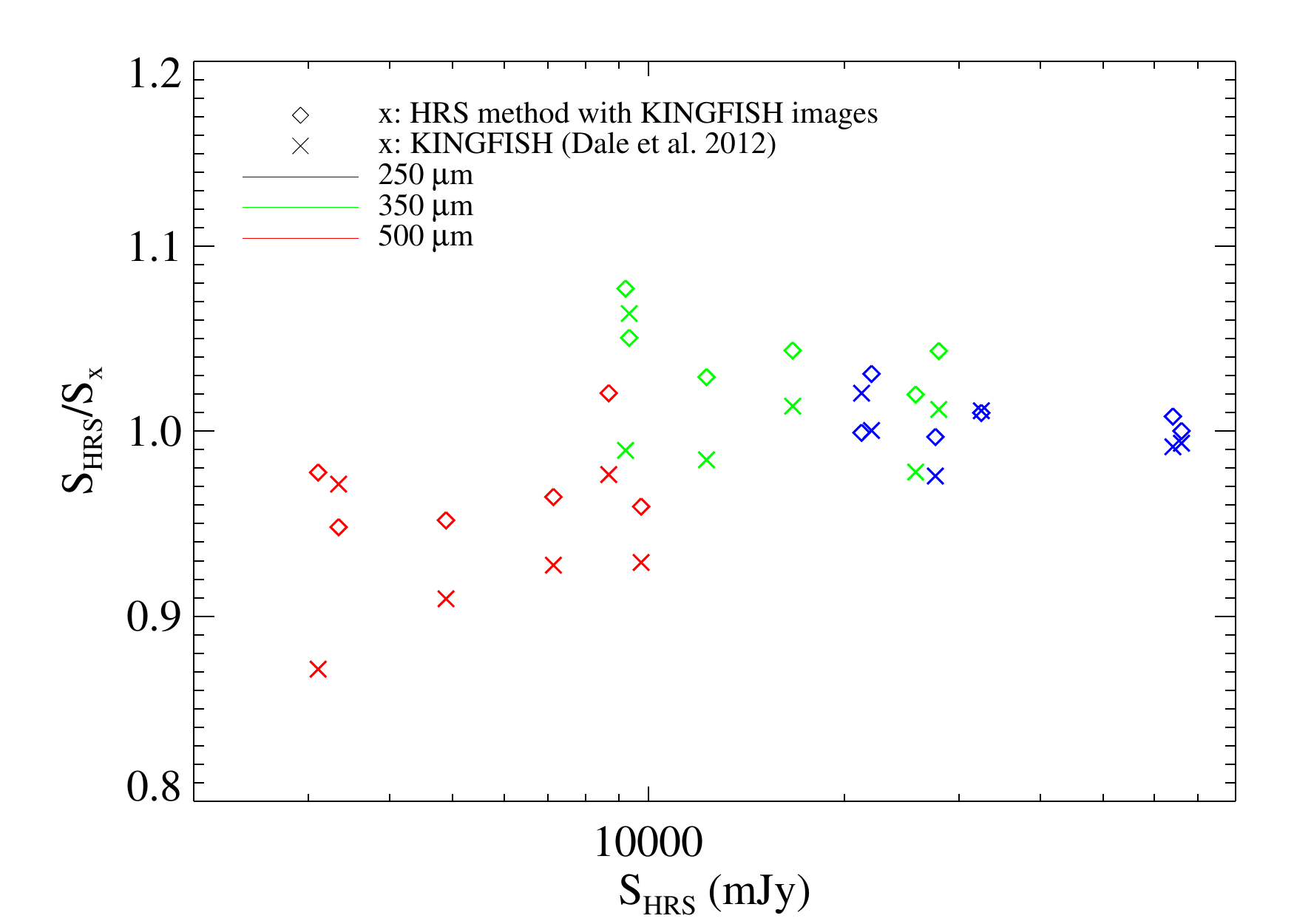}\\	
  	\caption{ \label{dalevsme}  Left panel: Comparison between KINGFISH \citep{Dale12} and HRS flux densities of six common galaxies: NGC~4254 (HRS~102), NGC~4321 (HRS~122), NGC~4536 (HRS~205), NGC~4569 (HRS~217), NGC~4579 (HRS~220) and NGC~4725 (HRS~263) in the three SPIRE bands. Blue diamonds, green triangles and red squares are for the 250, 350 and 500~$\mu$m measurements respectively. The dashed lines give the linear fits. Right panel: the flux density ratio obtained in this work to those given by \cite{Dale12} (crosses), as well as to those that we have extracted using our own procedure from the public KINGFISH images (diamonds).}
	\end{figure*}

	Flux densities are compared in Figure~\ref{dalevsme} (left panel).
 	Mean HRS to KINGFISH flux density ratios are $1.00\pm0.02$, $1.01\pm0.03$ and $0.93\pm 0.04$ at 250, 350, and 500~$\mu$m.
	At  250 and 350~$\mu$m the results are in very good agreement but not at 500~$\mu$m.
	This 7\% discrepancy can be due to the differences in the data reduction and map-making or in the flux extraction procedures.
	To understand if its origin comes from the method used to perform the photometry, we apply our flux extraction technique on the public KINGFISH images.
	On Figure~\ref{dalevsme} (right panel), we show the flux density ratio of the measurements obtained in this work to those given by \cite{Dale12} (crosses), as well as to those that we have extracted using our own procedure on the public KINGFISH images (diamonds).
	At 250~$\mu$m, all sets of data are consistent.
	At 350~$\mu$m, the flux densities measured on KINGFISH images with our procedure are $\sim4\%$ higher than ours.
	As the same method to extract fluxes is employed, this difference is due to the data reduction and map-making procedure applied by the SAG2 and the KINGFISH team.
	We note also a $\sim4\%$ systematic difference between \cite{Dale12} measurements and ours on the same images.
	This is due to the different flux extraction methods.
	At 500~$\mu$m, the flux densities measured on KINGFISH images with our procedure are $\sim3\%$ lower than our measurements, this discrepancy comes from the different production of the images.
	As at 350~$\mu$m, we note the $\sim4\%$ systematic error due to the different methods used to perform the photometry.
	The discrepancies observed between HRS and KINGFISH flux densities is thus a combination between differences in the production of the images and differences in the way the photometry is performed.

	\subsection*{Comparison with \textit{Planck}}
	
	We also compare our measurements with the \textit{Planck} Early Release Compact Source Catalog (ERCSC) of the \textit{Planck} Collaboration \citep{Planckcatalogue}.
	The catalogue contains flux densities derived from several method.
	To be consistent with this work, we use the measurements determined from aperture photometry.
	Cross-matching the two catalogues, we find 155 galaxies in common at 350~$\mu$m and 76 galaxies in common at 550~$\mu$m.
	The \textit{Planck} FWHM are $4.23'$ at 350~$\mu$m and $4.47'$ at 550~$\mu$m.
	The photometry on the \textit{Planck} compact sources was carried out using the FWHM of the band as the radius of the aperture.
	After visually inspecting each HRS galaxy with a corresponding \textit{Planck} source, we excluded 11 sources because the \textit{Planck} measurements may have potentially included other bright sources that lie within ~5\arcmin of the galaxies.	
	
\begin{figure*}
 	\includegraphics[width=9cm]{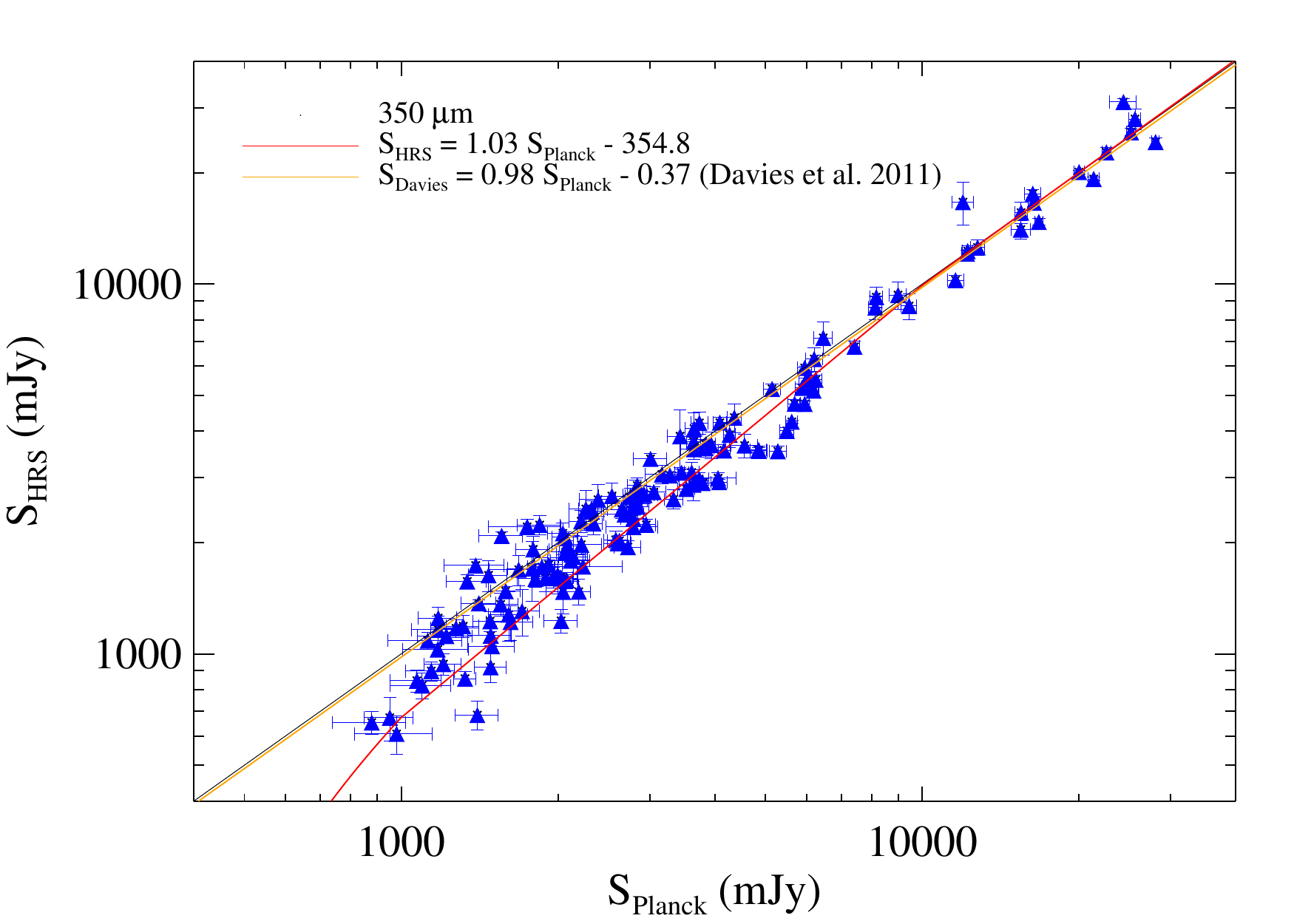}\hfill
 	\includegraphics[width=9cm]{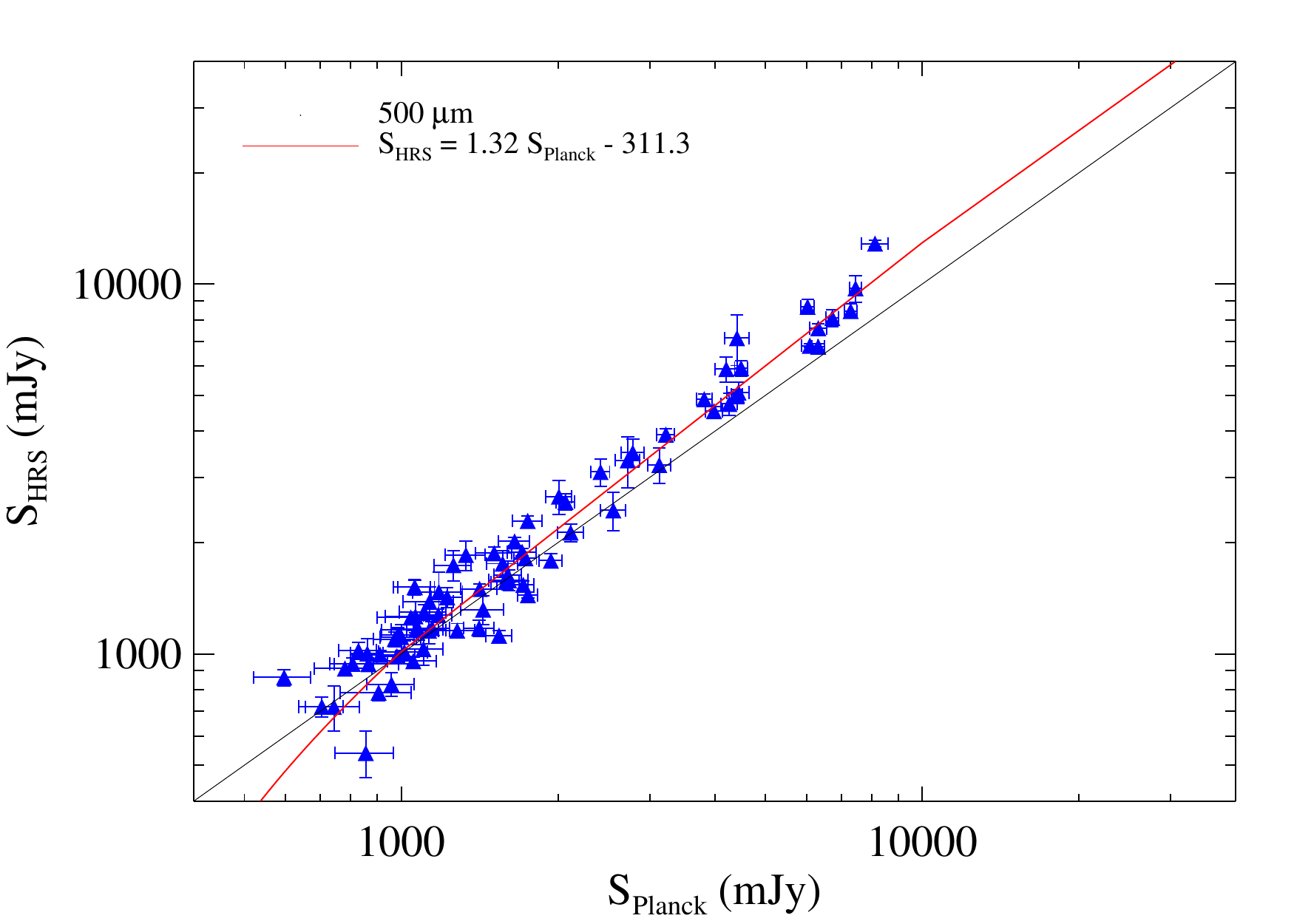}\\
	\includegraphics[width=9cm]{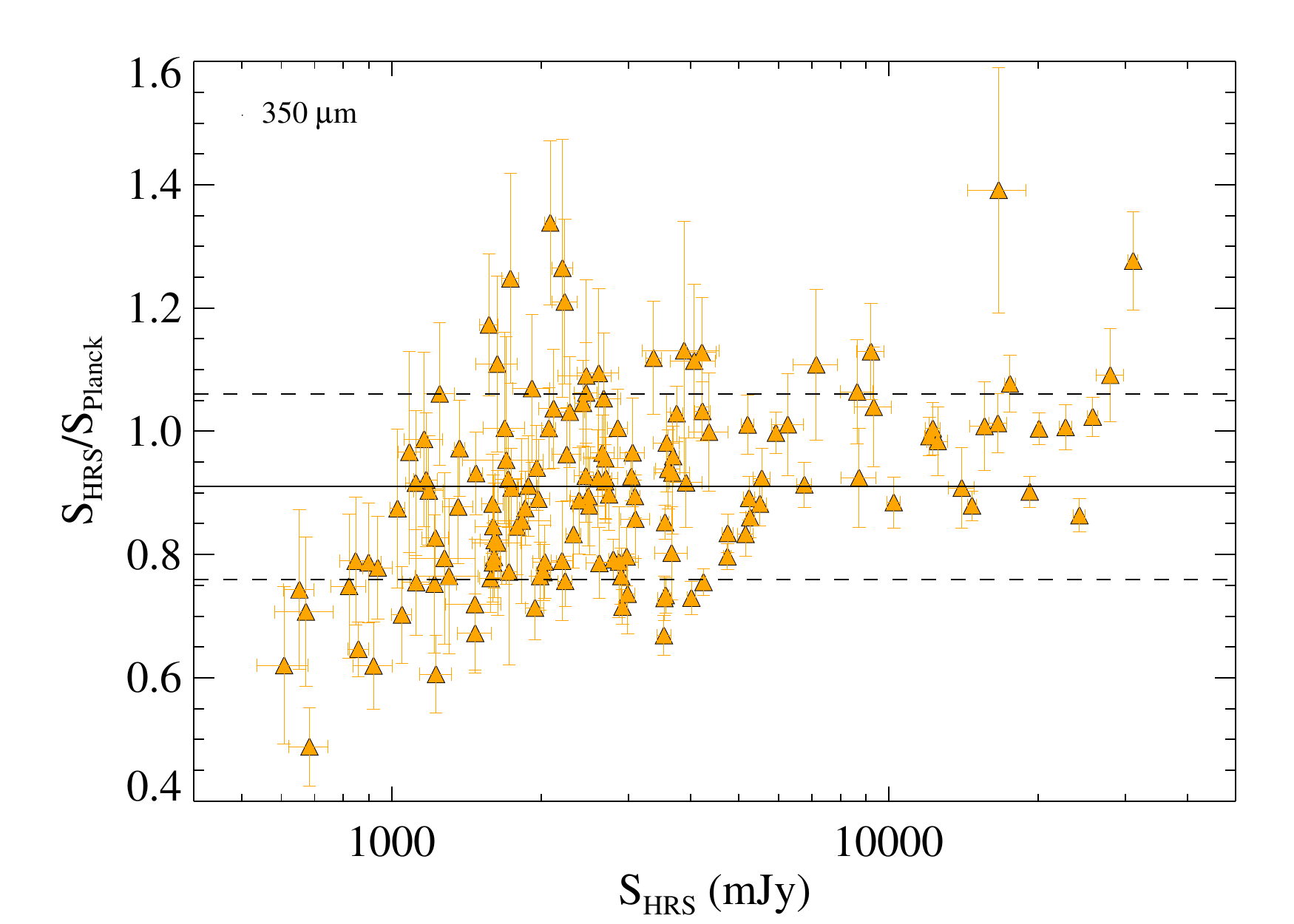}\hfill
 	\includegraphics[width=9cm]{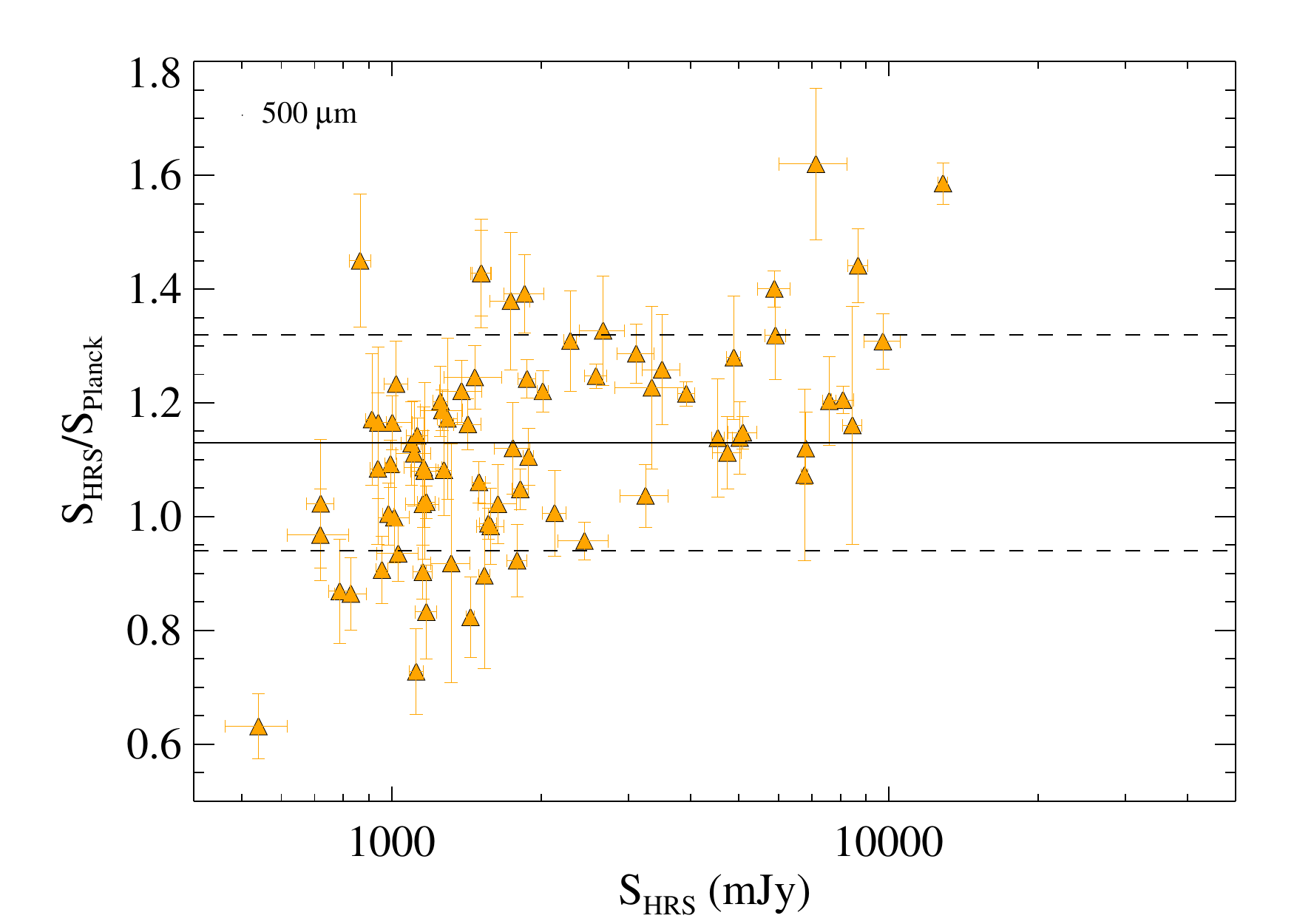}\\
	\includegraphics[width=9cm]{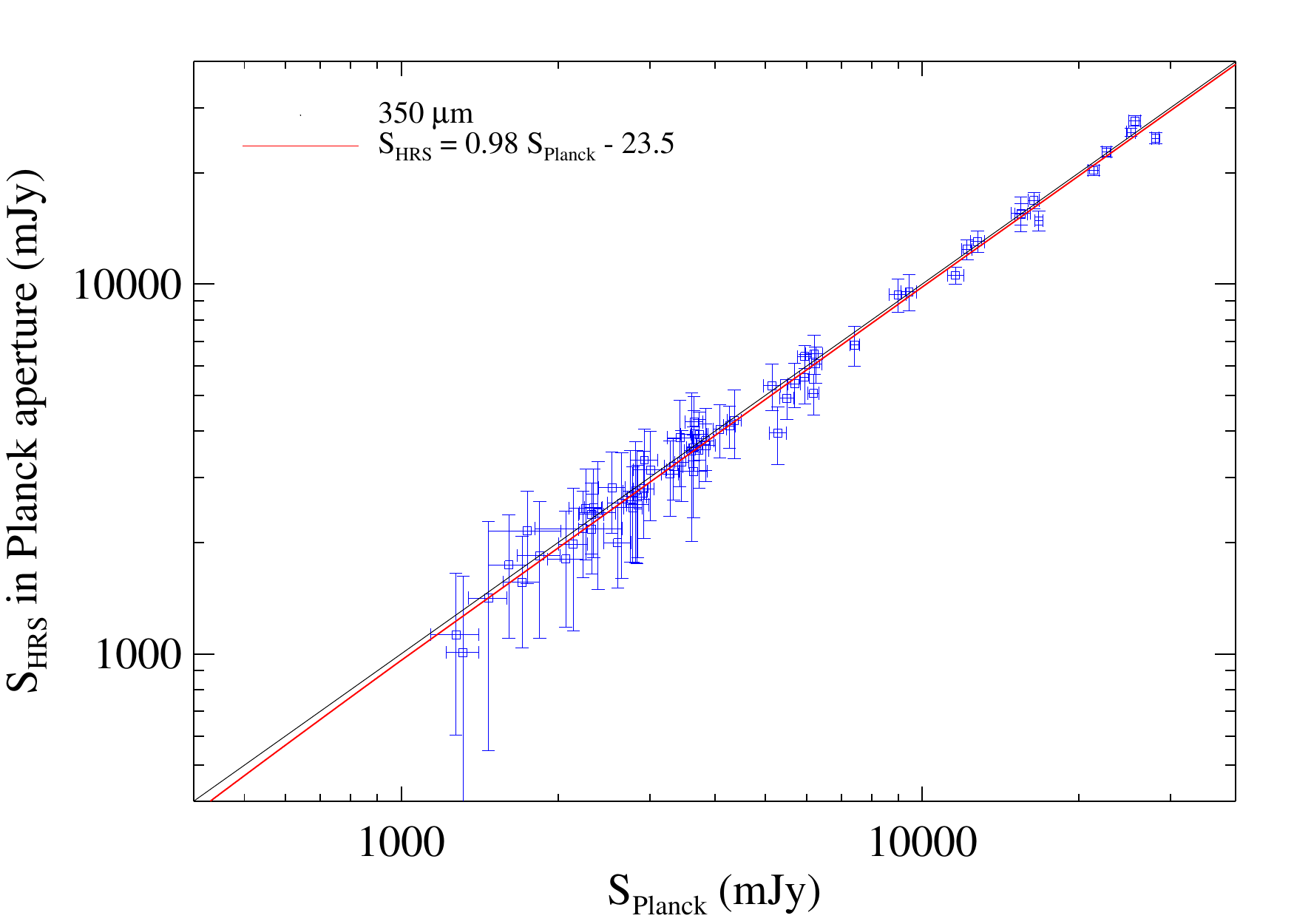}\hfill
 	\includegraphics[width=9cm]{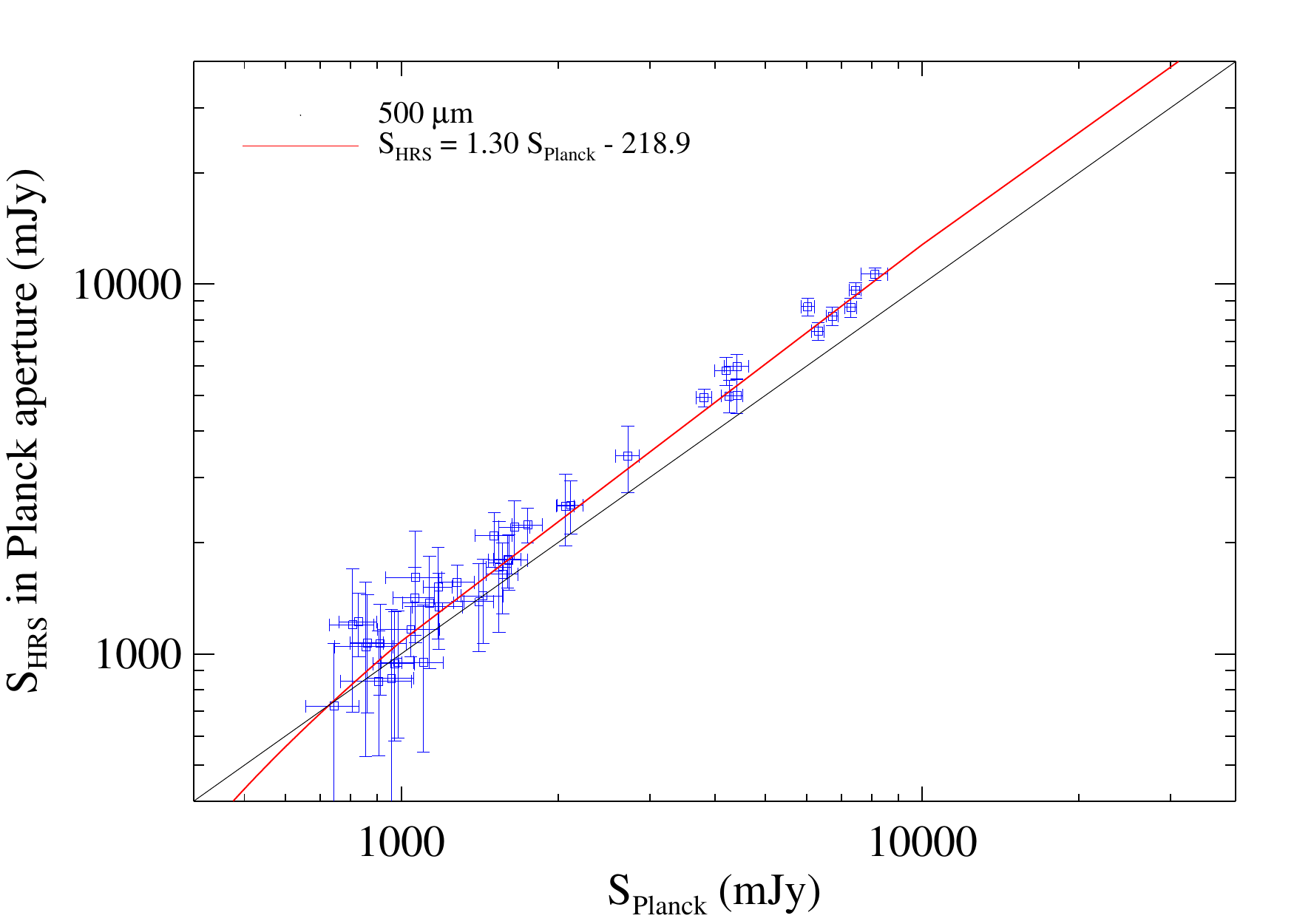}\\	
  	\caption{\label{Planckvsme} Upper panels: the HRS versus \textit{Planck} flux densities of 144 bright galaxies at 350~$\mu$m and 76 bright galaxies at 500~$\mu$m. The black lines are the one to one relationship in log scale. The red lines are the results of the linear fit; the orange line, on the 350~$\mu$m plot, is the result of \cite{Davies12} linear least squares fit. Middle panels: the HRS/\textit{Planck} flux density ratio versus the HRS flux densities at 350~$\mu$m and 500~$\mu$m. The black line corresponds to the mean ratio, the dashed lines correspond to the standard deviation of the ratios. Lower panels: \textit{Planck} flux densities versus HRS flux densities measured in the aperture used by the \textit{Planck} Consortium at 350~$\mu$m and 500~$\mu$m. }      
	\end{figure*}

\begin{figure*}
	\centering
 	\includegraphics[width=9cm]{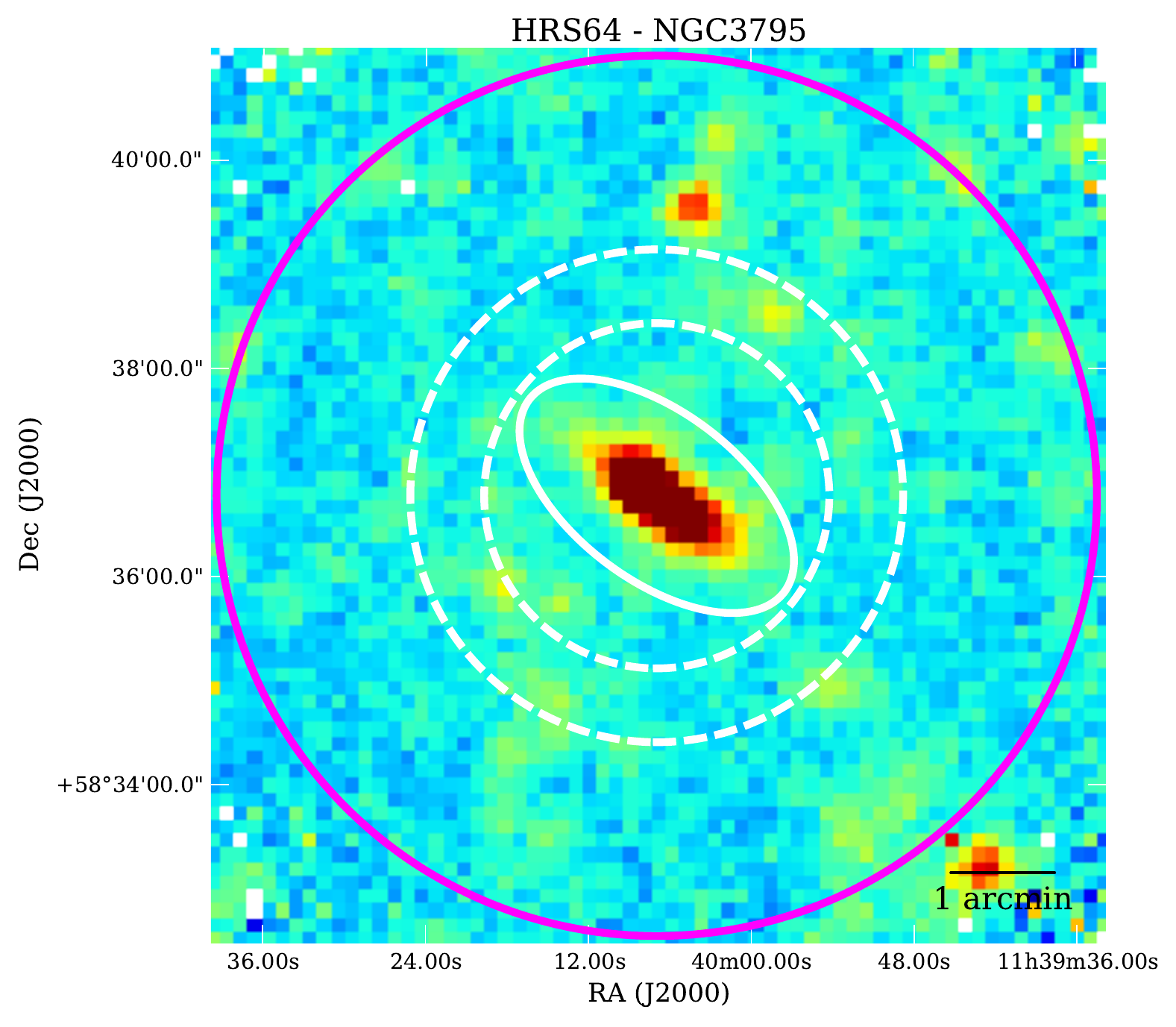}\hfill
 	\includegraphics[width=9cm]{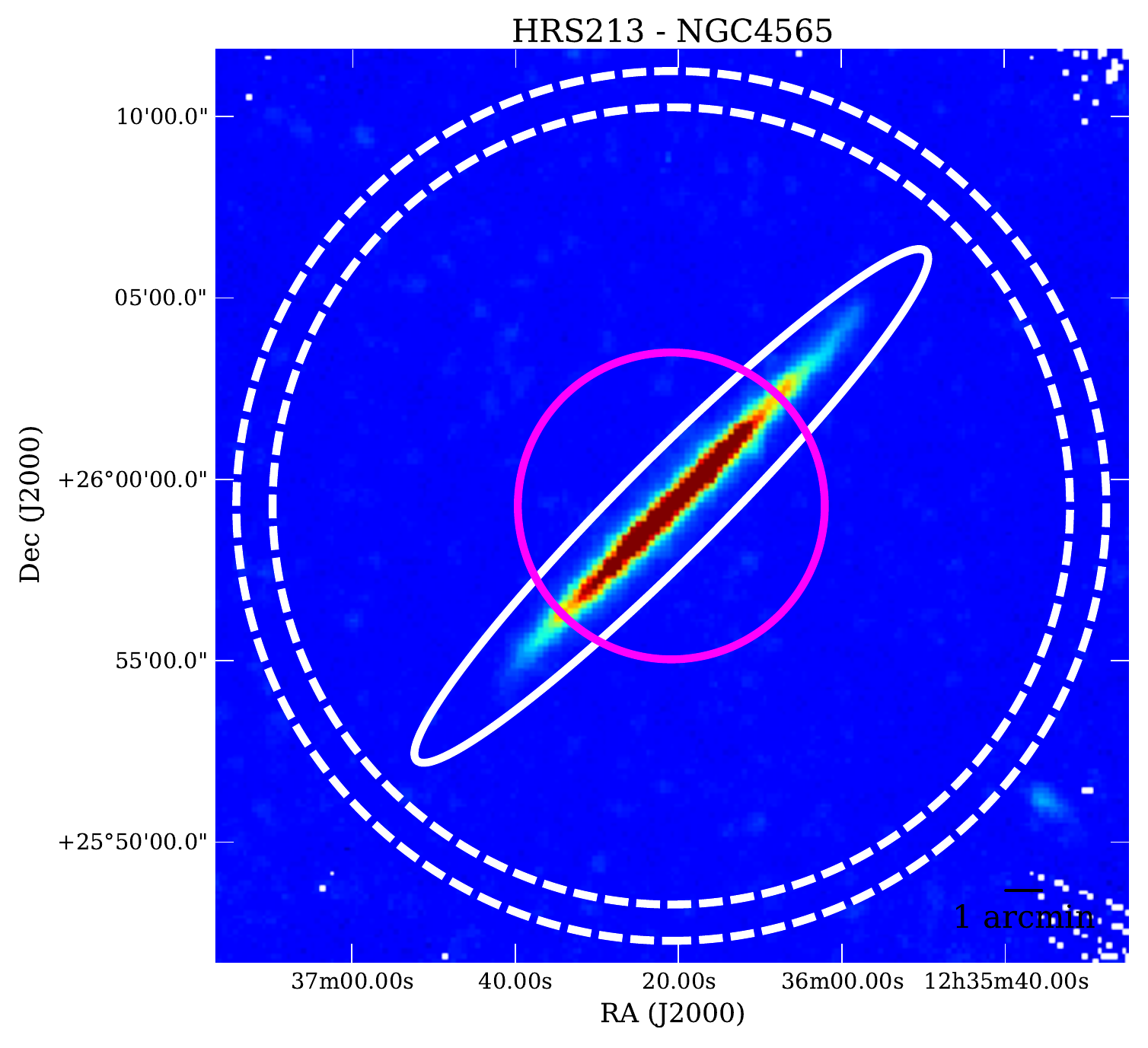}\\
 	\includegraphics[width=9cm]{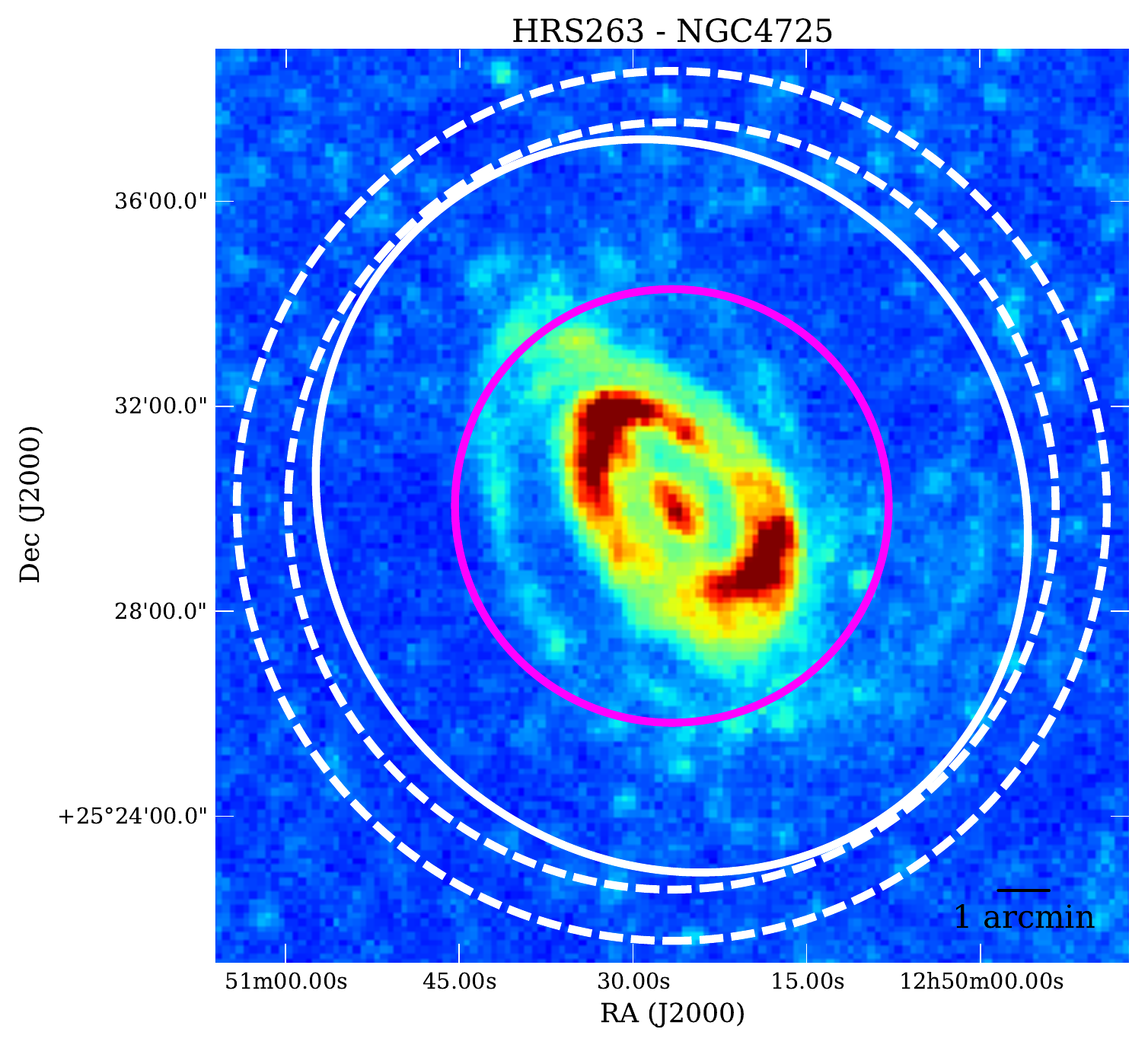}\hfill
 	\includegraphics[width=9cm]{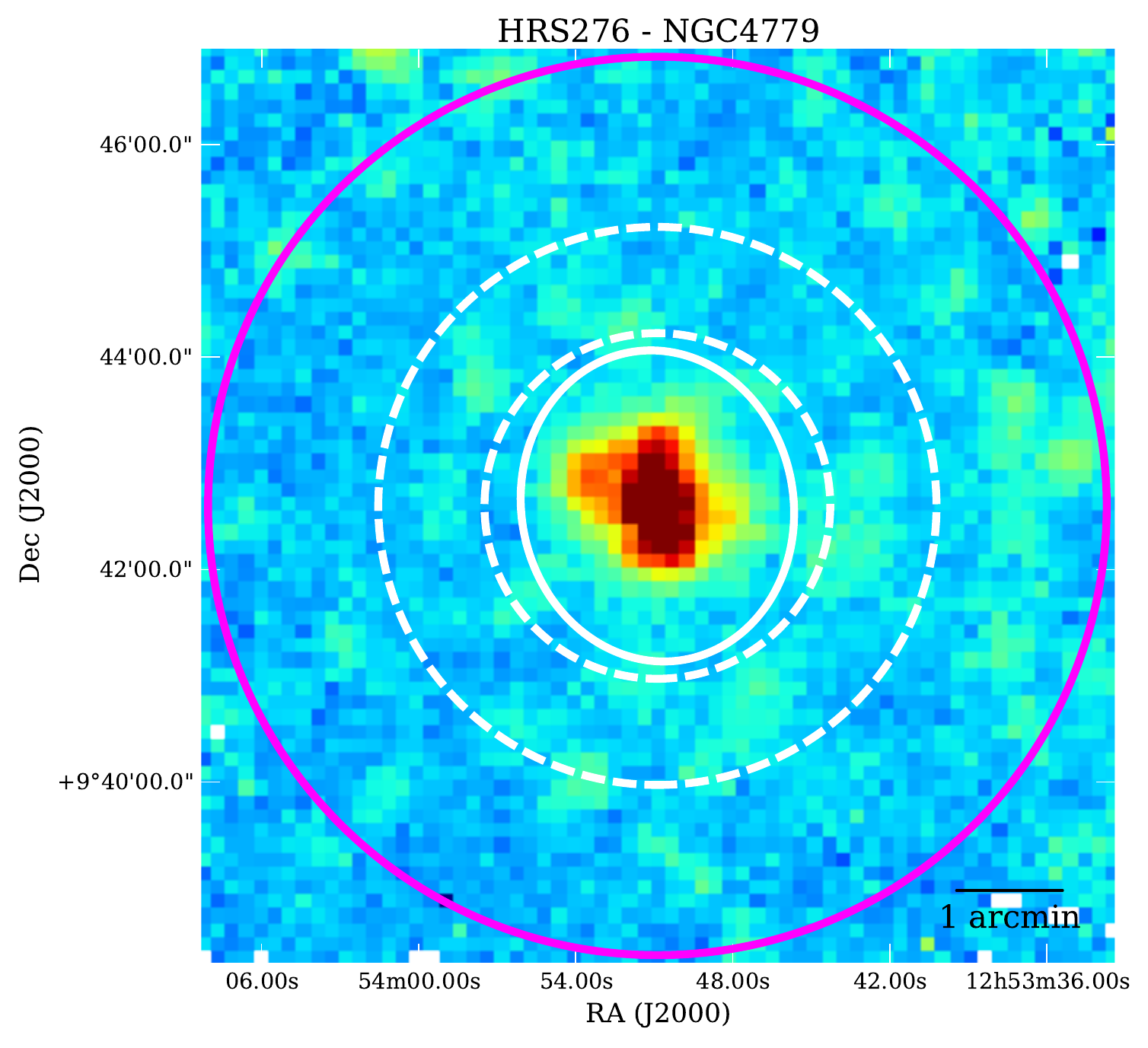}\\	
  	\caption{\label{Planckcomp} Comparison between \textit{Planck} and \textit{Herschel} aperture photometry on 350~$\mu$m SPIRE images, with in white, the apertures used for \textit{Herschel} photometry and in dashed white for the background estimation, in magenta the aperture used by the \textit{Planck} consortium on \textit{Planck} images.   }
	\end{figure*}

	The comparison between the data taken from the Planck catalogues and those presented here in Table~\ref{HRS_phot} is shown in Figure~\ref{Planckvsme}.
	Mean \textit{Herschel} to \textit{Planck} flux density ratios are 0.91 and 1.13 at 350 and 500~$\mu$m, respectively.
	At 500~$\mu$m, \textit{Herschel} flux densities are on average higher than those of \textit{Planck} at 550~$\mu$m.
	The discrepancy at 500~$\mu$m is in part due to the difference of wavelength. 
	If we assume a modified black body with a $\beta$ of 1.5, $T=$20K, and assuming the relative colour corrections (1.02 for the SPIRE data, 0.91 for the \textit{Planck} one, as indicated by the ERCSC Explanatory Supplement), we expect a flux density ratio of $S(500)_{SPIRE}/S(550)_{\textit{Planck}} = 1.15$. 
	Once correcting the \textit{Planck} data for this difference, the same ratio drops to 0.98 at 500~$\mu$m and 0.96 at 350~$\mu$m. 
	Colour corrections thus explain the mean differences between the two independent sets of data. 
	They do not however explain other systematic differences such as those related to aperture effects.
	The major differences between the \textit{Planck} measurements and ours are due to the different aperture sizes.
	Higher ratios correspond to galaxies that have a size larger than the \textit{Planck} 350~$\mu$m FWHM.	
	Lower ratios can be explained by the contribution of background sources, visible on \textit{Herschel} images, present in the \textit{Planck} beam (Figure~\ref{Planckcomp}). 
	At low flux densities, \textit{Planck} data are systematically higher than those of \textit{Herschel}, probably due to the important contamination of background sources (Figure~\ref{Planckvsme}).
	\cite{Davies12} also compared their results with those of the \textit{Planck} Collaboration at 350~$\mu$m. 
	The result of their linear fit is given in Figure~\ref{Planckvsme}, upper left panel (flux densities are not color corrected).

	To understand this strong systematic difference between \textit{Planck} measurements and ours, we carried out the photometric method used for the \textit{Planck} catalogue on HRS images of the galaxies in common but using the aperture and sky annulus defined by the \textit{Planck} consortium.
	A circular aperture with a radius of 4.23\arcmin (4.47\arcmin) at 350~$\mu$m (500~$\mu$m) is used, and the background region is defined as a circular annulus with an inner radius of 4.23\arcmin (4.47\arcmin) at 350~$\mu$m (500~$\mu$m) and outer radius of 2$\times$4.23\arcmin (2$\times$4.47\arcmin) at 350~$\mu$m (500~$\mu$m).
	This is a rough approximation, and a more precise work would require a convolution of the \textit{Herschel} images to the resolution of \textit{Planck} which is beyond the purpose of the present paper.	
	We rejected flux densities of galaxies with sizes bigger than the \textit{Planck} beam, or faint galaxies contaminated by strong background sources within the \textit{Planck} beam.
 	Figure~\ref{Planckvsme} (lower panels) shows the \textit{Planck} photometry versus HRS photometry measured in \textit{Planck} apertures only for galaxies with good flux density measurements, the associated flux densities are in Table~\ref{4galpl}.
	The results of the linear least squares fits show the consistency of the two sets of measurements.
	The HRS photometry in the beam of \textit{Planck} is in good agreement with the flux densities of the \textit{Planck} catalogue.
	Furthermore, the calibration error of \textit{Planck} at 350 and 550~$\mu$m \textit{Planck} is 7\% \citep{PlanckHFICoreTeam}, associated with the calibration error of \textit{Herschel}, we obtain 10\%. 
	We can conclude that, despite differences due to the choice of  the aperture, our measurements are consistent with the \textit{Planck} data.
	However, because of the various photometry issues but particularly the source blending issues, \textit{Herschel} measurements should be used instead of \textit{Planck} measurements for these galaxies whenever possible.
	
	\begin{table}
	\centering
	\caption{Flux densities from different sources for the 4 galaxies shown in Figure~\ref{Planckcomp}, at 350~$\mu$m.}
	\begin{tabular}{l c c c}
  	\hline\hline
  	  					&HRS  & \textit{Planck}  & HRS in \textit{Planck} beam \\
						& mJy	&	mJy		 &	mJy \\
  	\hline 
  	HRS~64			& $683 \pm 62$		& $1400 \pm 132$		& $1372 \pm 810$ \\ 
  	HRS~213 			& $31070 \pm 687$	& $24342 \pm 1431$ 	&  $25504 \pm 922$\\ 
	HRS~263 			& $16657 \pm 2242$	& $11974 \pm 574$	&  $14144 \pm 815$ \\
	HRS~276 			& $1229 \pm 90$		& $2028 \pm 147$		&  $1904 \pm 809$  \\  
	\hline
	\label{4galpl}
	\end{tabular}
	\end{table}

\section{Data access}
The table is available on the SAG2 \textit{Herschel} Database in Marseille (HeDaM; Roehlly et al. in preparation) at \url{http://hedam.oamp.fr/HRS/}.
An electronic version of the catalogue and a README file can be downloaded there.
The README describes how the photometry is performed for both extended and point-like sources. 
Through this database, we plan in the next future to give access to the community to all \textit{Herschel} and ancillary data of the HRS galaxies.
\section{Conclusion}

We present the flux densities of the 323 galaxies of the \textit{Herschel} Reference Survey in the three SPIRE bands.
For extended galaxies, aperture photometry on elliptical regions is performed using the "Funcnts" DS9/Funtools task.
The background contribution is estimated calculating the mean value of the pixels within a concentric circular annulus.
A different technique is used  for point-like sources, where a PSF fitting is directly performed on timeline data.
We compare our results with those of \cite{Davies12}, KINGFISH \citep{Dale12} and the \textit{Planck} Early Science Compact Source Catalog \citep{Planckcatalogue}.
Our measurements and those of \cite{Davies12} and \cite{Dale12} are consistent. 
Despite the different size of PSF between SPIRE and \textit{Planck}, our flux densities and those of the \textit{Planck} Consortium are in a good agreement. 
The catalogue is publicly available on the HeDaM database.
\begin{acknowledgements}
We thank the referee for precious comments and suggestions which helped improving the 
quality of the manuscript.
LC thanks Daniel Dale for enlightening discussions about the photometry of extended galaxies.
LC although thanks Samuel Boissier and S\'ebastien Heinis for useful discussions. 
AB thanks the ESO visiting program committee for inviting him at the Garching headquarters for a two months staying.
SB, SdiSA and CP acknowledge financial support by ASI through the ASI-INAF grants I/016/07/0 and I/009/10/0.
SPIRE has been developed by a consortium of institutes led
by Cardiff Univ. (UK) and including Univ. Lethbridge (Canada); NAOC (China);
CEA, LAM (France); IFSI, Univ. Padua (Italy); IAC (Spain); Stockholm
Observatory (Sweden); Imperial College London, RAL, UCL-MSSL, UKATC,
Univ. Sussex (UK); Caltech, JPL, NHS C, Univ. Colorado (USA). This development
has been supported by national funding agencies: CSA (Canada); NAOC
(China); CEA, CNES, CNRS (France); ASI (Italy); MCINN (Spain); SNSB
(Sweden); STFC, UKSA (UK); and NASA (USA). This research has made
use of the NASA/IPAC ExtraGalactic Database (NED) which is operated by
the Jet Propulsion Laboratory, California Institute of Technology, under contract
with the National Aeronautics and Space Administration. The research
leading to these results has received funding from the European CommunityÕs
Seventh Framework Programme (/FP7/2007-2013/) under grant agreement
No 229517. This research has made use of the NASA/IPAC ExtraGalactic
Database (NED) which is operated by the Jet Propulsion Laboratory, California
Institute of Technology, under contract with the National Aeronautics and Space
Administration and of the GOLDMine database (http://goldmine.mib.infn.it/).
The Dark Cosmology Centre is funded by the Danish National Research Foundation.

\end{acknowledgements}


\bibliographystyle{aa}
\bibliography{hrs_photometry}

\appendix
\section{Tables}	
	\tiny
\onecolumn{
\begin{longtable}{c c c c c c c c c c c c c c }
\caption{\label{apertures} Apertures used for the on source and sky background estimates (see Section~\ref{ext}).} \\
\hline
\hline
HRS  & CGCG   &  VCC   & UGC  &  NGC  &  IC   & RA & Dec &  $a_{IR}$ & $b_{IR}$ & $PA$ &  $r_{in}^{bck}$  &  $r_{out}^{bck}$  \\
    &      &    &     &      &   &   &   & (arcsec)               &  (arcsec)          &  (degree)      &   (arcsec)               &  (arcsec)          \\
\hline
 \endfirsthead
\caption{continued.}\\  
\hline
\hline
HRS  & CGCG   &  VCC   & UGC  &  NGC  &  IC   & RA & Dec &  $a_{IR}$ & $b_{IR}$ & $PA$ &  $r_{in}^{bck}$  &  $r_{out}^{bck}$  \\
   &      &    &     &      &   &   &   & (arcsec)               &  (arcsec)          &  (degree)      &   (arcsec)               &  (arcsec)          \\
\hline
\endhead
\hline
\endfoot
       1    &    123035    &         -    &         -    &         -    &         -    &         154.415    &         22.8100    &        47    &        40    &       170    &        46    &       106    \\
       2    &    124004    &         -    &      5588    &         -    &         -    &         155.238    &         25.3650    &        46    &        45    &        40    &        62    &       103    \\
       3    &     94026    &         -    &      5617    &      3226    &         -    &         155.863    &         19.8980    &        39    &        38    &        15    &       300    &       354    \\
       4    &     94028    &         -    &      5620    &      3227    &         -    &         155.877    &         19.8650    &       117    &        97    &       155    &       249    &       309    \\
       5    &     94052    &         -    &         -    &         -    &       610    &         156.618    &         20.2280    &        80    &        47    &        28    &        86    &       146    \\
       6    &    154016    &         -    &      5662    &      3245    &         -    &         156.755    &         28.6390    &       125    &        45    &       150    &       153    &       213    \\
       7    &    154017    &         -    &      5663    &      3245    &         -    &         156.827    &         28.5070    &        20    &        20    &         -    &       150    &       210    \\
       8    &    154020    &         -    &      5685    &      3254    &         -    &         157.333    &         29.4910    &       210    &        66    &        46    &       232    &       292    \\
\hline
\end{longtable}
}
\twocolumn

	\onecolumn{
\begin{center}
\begin{longtable}{p{0.2cm} p{0.1cm} p{1.9cm} p{0.1cm} p{1.9cm} p{0.1cm} p{1.9cm} p{0.3cm} p{0.3cm} p{0.3cm} p{0.4cm} p{0.4cm} p{0.4cm} p{0.4cm} p{0.4cm} p{0.4cm} p{0.4cm} p{0.4cm} p{0.4cm} p{0.4cm} p{0.4cm} p{0.4cm} p{0.4cm}}
\caption{\label{HRS_phot} Flux densities, numbers of pixels encompassed within the apertures, and errors of the HRS galaxies at 250, 350 and 500~$\mu$m, see Section~\ref{fluxdens}.} \\
\hline
\hline
\tiny
HRS & $f_{\rm 250}$ & $S_{\rm 250}$\tablefootmark{1} & $f_{\rm 350}$ & $S_{\rm 350}$\tablefootmark{1} & $f_{\rm 500}$ & $S_{\rm 500}$\tablefootmark{1} & $N_{\rm 250}$ & $N_{\rm 350}$ & $N_{\rm 500}$ & $err^{\rm 250}_{\rm inst}$ & $err^{\rm 350}_{\rm inst}$ & $err^{\rm 500}_{\rm inst}$ & $err^{\rm 250}_{\rm conf}$ & $err^{\rm 350}_{\rm conf}$ & $err^{\rm 500}_{\rm conf}$ & $err^{\rm 250}_{\rm sky}$ & $err^{\rm 350}_{\rm sky}$ & $err^{\rm 500}_{\rm sky}$\\
    &  &  (mJy) & & (mJy)&   &   (mJy) &   & & & (mJy) &  (mJy)  &  (mJy) & (mJy) & (mJy)  & (mJy)  & (mJy) & (mJy) & (mJy)     \\
\hline
 \endfirsthead
\caption{continued.}\\  
\hline
\hline
HRS & $f_{\rm 250}$ & $S_{\rm 250}$\tablefootmark{1} & $f_{\rm 350}$ & $S_{\rm 350}$\tablefootmark{1} & $f_{\rm 500}$ & $S_{\rm 500}$\tablefootmark{1} & $N_{\rm 250}$ & $N_{\rm 350}$ & $N_{\rm 500}$ & $err^{\rm 250}_{\rm inst}$ & $err^{\rm 350}_{\rm inst}$ & $err^{\rm 500}_{\rm inst}$ & $err^{\rm 250}_{\rm conf}$ & $err^{\rm 350}_{\rm conf}$ & $err^{\rm 500}_{\rm conf}$ & $err^{\rm 250}_{\rm sky}$ & $err^{\rm 350}_{\rm sky}$ & $err^{\rm 500}_{\rm sky}$\\
    & &  (mJy) & & (mJy)&   &   (mJy) &   & & & (mJy) &  (mJy)  &  (mJy) & (mJy) & (mJy)  & (mJy)  & (mJy) & (mJy) & (mJy)     \\
\hline
\endhead
\hline
\endfoot
       1   & $        1 $ & $          498.3   \pm           37.2 $ & $        1 $ & $          210.8   \pm           29.1 $ & $        1 $ & $           79.3   \pm           17.5  $ &       163    &        92    &        41   & $            6.7 $ & $            5.0 $ & $            4.0 $ & $           21.2 $ & $           17.4 $ & $           12.7 $ & $           29.8 $ & $           22.8 $ & $           11.3 $  \\
       2   & $        1 $ & $         1163.0   \pm           58.5 $ & $        1 $ & $          482.7   \pm           36.2 $ & $        1 $ & $          153.0   \pm           18.2  $ &       185    &       102    &        45   & $            7.9 $ & $            5.6 $ & $            4.2 $ & $           22.6 $ & $           18.3 $ & $           13.3 $ & $           53.3 $ & $           30.8 $ & $           11.7 $  \\
       3   & $        1 $ & $          580.9   \pm           33.8 $ & $        1 $ & $          241.8   \pm           23.9 $ & $        1 $ & $           79.9   \pm           13.8  $ &       135    &        75    &        32   & $            3.5 $ & $            2.5 $ & $            2.0 $ & $           19.3 $ & $           15.7 $ & $           11.2 $ & $           27.5 $ & $           17.9 $ & $            7.7 $  \\
       4   & $        1 $ & $        12674.1   \pm          209.9 $ & $        1 $ & $         5240.7   \pm          140.6 $ & $        1 $ & $         1815.2   \pm           68.2  $ &       993    &       560    &       251   & $           16.2 $ & $            8.7 $ & $            6.2 $ & $           52.4 $ & $           42.9 $ & $           31.5 $ & $          202.6 $ & $          133.6 $ & $           60.2 $  \\
\hline

\end{longtable}
\end{center}
\tablefoot{Errors in this table do not contain the 7$\%$ calibration errors. Flux densities in this table do not contain colour corrections.}\\
\tablefoottext{a}{Presence of a companion galaxy; flux densities are overestimated.}\\
\tablefoottext{b}{Presence of a background source that cannot be separated at 250, 350 and 500~$\mu$m; flux densities are overestimated.}\\
\tablefoottext{c}{Presence of a background source that cannot be separated at 350 and 500~$\mu$m; flux densities are overestimated.}	\\	
\tablefoottext{d}{Presence of a strong cirrus.}\\
\tablefoottext{e}{The source is considered point like; however in the PLW band (500~$\mu$m), even if there is a detection, the emission of the galaxy is dominated by a background source; the 500~$\mu$m flux density is thus an upper limit. }\\
\tablefoottext{f}{Particular aperture adapted to take into account M86's structures \citep{Gomez10,Cortese10a}. }\\
\tablefoottext{g}{Aperture is off-centered to match the particular shape of NGC~4438 \citep{Cortese10a}.}\\
\tablefoottext{h}{Flux densities in SPIRE bands dominated by synchrotron emission \citep{Baes10,Boselli10b}}\\
\tablefoottext{i}{Presence of the companion NGC~4496B; flux densities are overestimated.}\\
\tablefoottext{j}{NGC~4567 and NGC~4568, the two galaxies are overlapping; flux densities are overstimated.}\\
\tablefoottext{k}{Presence of a background source that cannot be separated at 500~$\mu$m; flux are overestimated.}\\
}
\twocolumn 
	\begin{figure*}
  \centering
  \subfloat{\includegraphics[width=\textwidth]{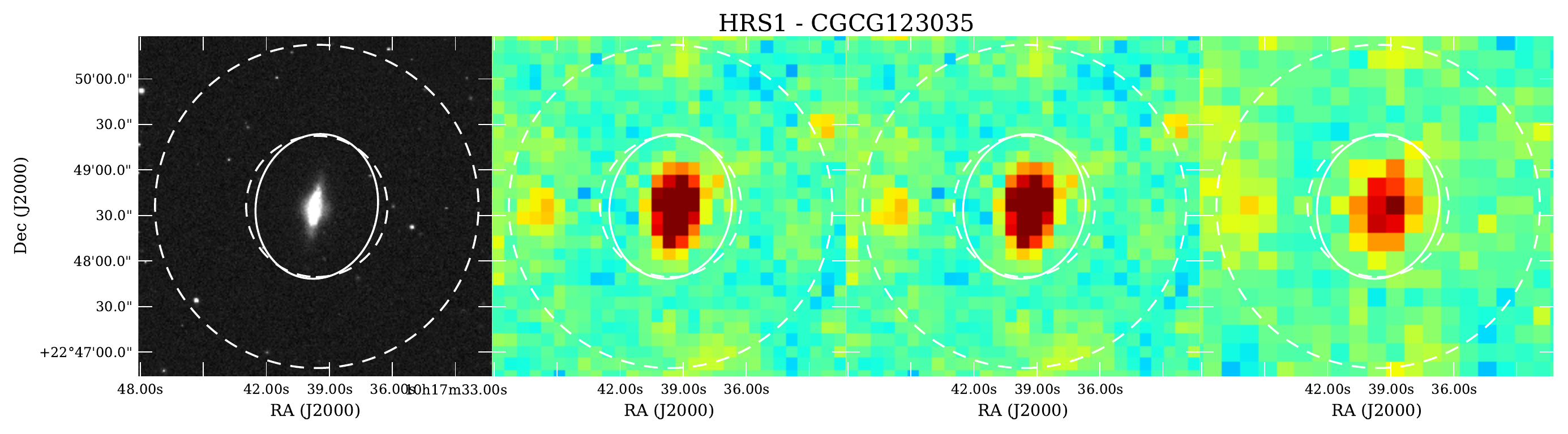}}\\
  \subfloat{\includegraphics[width=\textwidth]{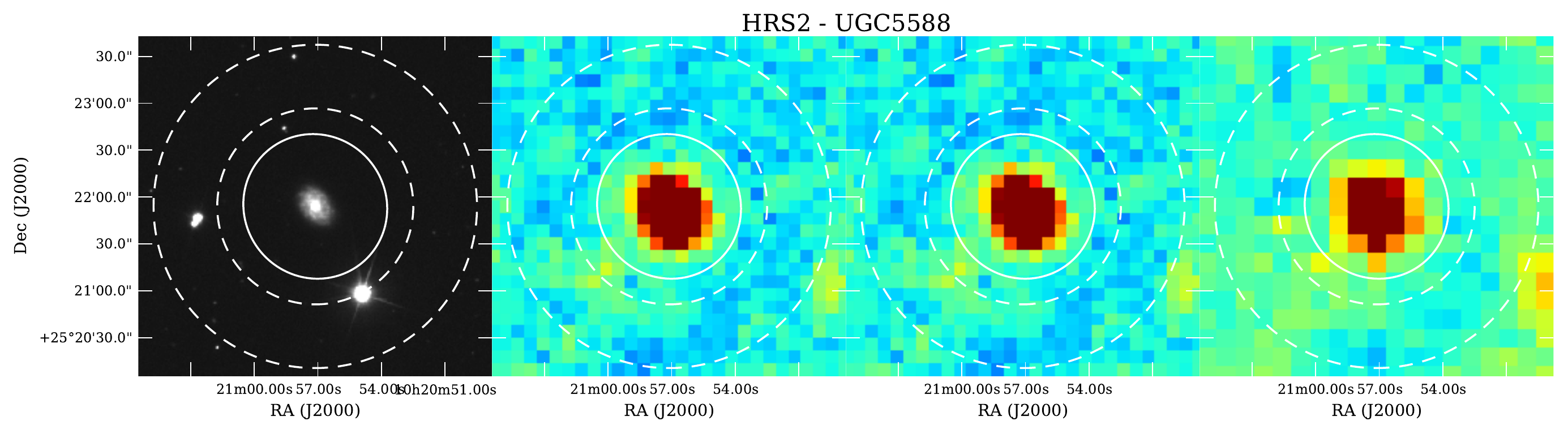}}\\
  \subfloat{\includegraphics[width=\textwidth]{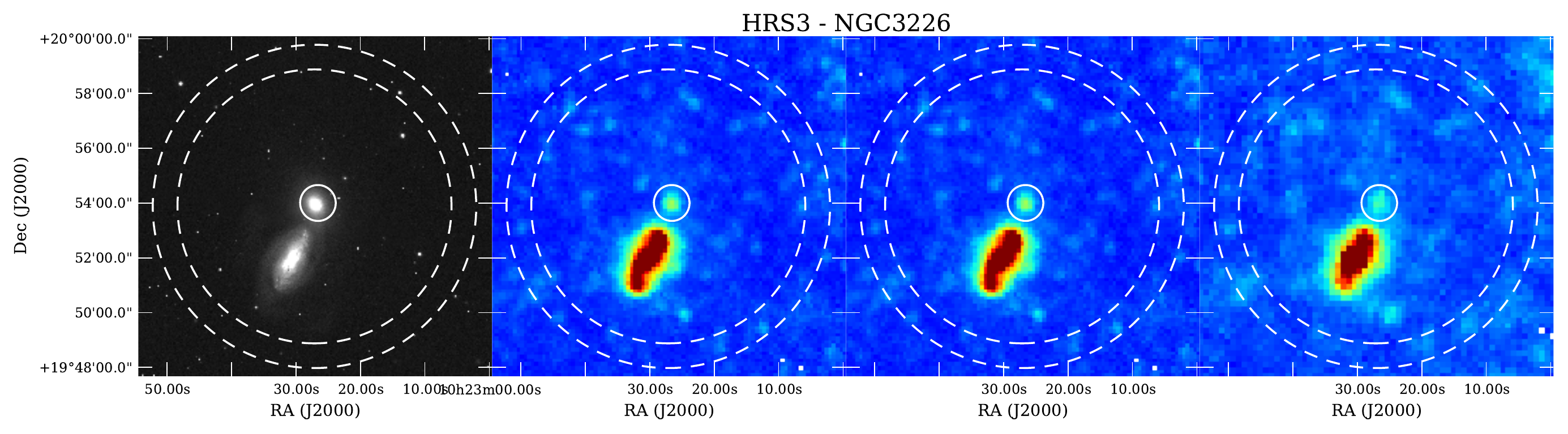}}\\
  \subfloat{\includegraphics[width=\textwidth]{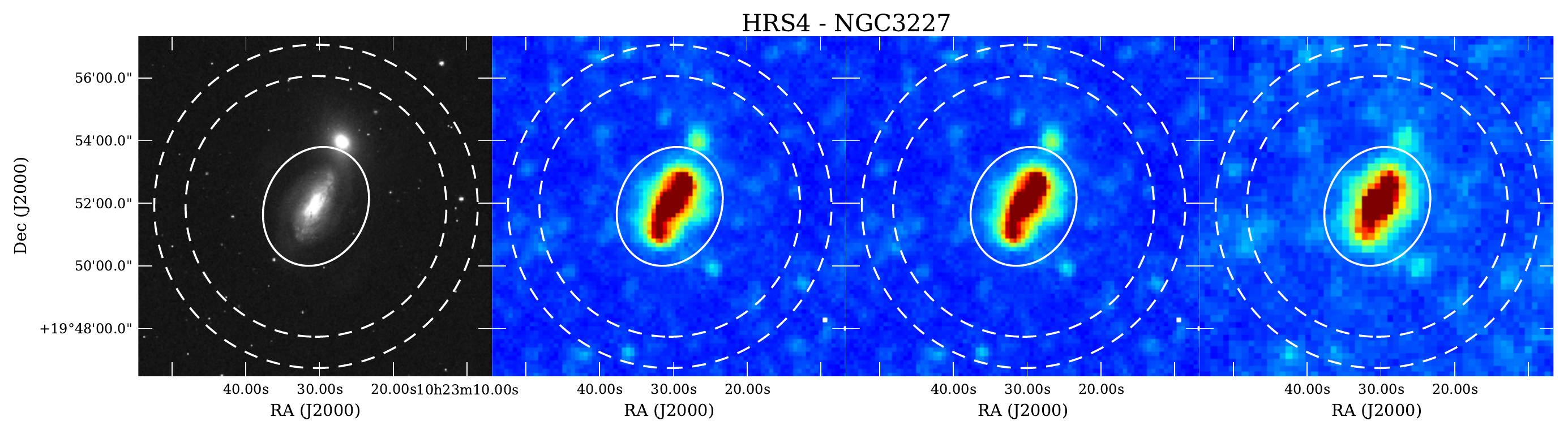}}\\
   \caption{\label{allfigures} Images of 4 HRS galaxies. From left to right: SDSS \textit{r'} band image, 250, 350, and 500~$\mu$m \textit{Herschel} images. The aperture used for the photometry is indicated by the solid line and the annulus, where the background is estimated, is indicated in dashed lines. The images of all the HRS galaxies are available on Hedam: http://hedam/HRS/index.php. }
\end{figure*}
\clearpage

\end{document}